\documentclass[aps,pre,twocolumn,showpacs]{revtex4-2}
\usepackage{graphicx}
\usepackage{dcolumn}
\usepackage{bm}
\usepackage{here}
\usepackage{color}
\usepackage{braket}
\usepackage{ascmac}

\begin{document}

\newcommand{\vc}[1]{\mbox{\boldmath $#1$}}
\newcommand{\fracd}[2]{\frac{\displaystyle #1}{\displaystyle #2}}
\newcommand{\red}[1]{\textcolor{red}{#1}}
\newcommand{\blue}[1]{\textcolor{blue}{#1}}
\newcommand{\green}[1]{\textcolor{green}{#1}}



\def\ni{\noindent}
\def\nn{\nonumber}
\def\bH{\begin{Huge}}
\def\eH{\end{Huge}}
\def\bL{\begin{Large}}
\def\eL{\end{Large}}
\def\bl{\begin{large}}
\def\el{\end{large}}
\def\beq{\begin{eqnarray}}
\def\eeq{\end{eqnarray}}
\def\beqnn{\begin{eqnarray*}}
\def\eeqnn{\end{eqnarray*}}

\def\bit{\begin{itemize}}
\def\eit{\end{itemize}}
\def\bsc{\begin{screen}}
\def\esc{\end{screen}}

\def\eps{\epsilon}
\def\th{\theta}
\def\del{\delta}
\def\omg{\omega}

\def\e{{\rm e}}
\def\exp{{\rm exp}}
\def\arg{{\rm arg}}
\def\Im{{\rm Im}}
\def\Re{{\rm Re}}

\def\sup{\supset}
\def\sub{\subset}
\def\a{\cap}
\def\u{\cup}
\def\bks{\backslash}

\def\ovl{\overline}
\def\unl{\underline}

\def\rar{\rightarrow}
\def\Rar{\Rightarrow}
\def\lar{\leftarrow}
\def\Lar{\Leftarrow}
\def\bar{\leftrightarrow}
\def\Bar{\Leftrightarrow}

\def\pr{\partial}

\def\>{\rangle} 
\def\<{\langle} 
\def\RR {\rangle\!\rangle} 
\def\LL {\langle\!\langle} 
\def\const{{\rm const.}}

\def\e{{\rm e}}

\def\Bstar{\bL $\star$ \eL}

\def\etath{\eta_{th}}
\def\irrev{{\mathcal R}}
\def\e{{\rm e}}
\def\noise{n}
\def\hatp{\hat{p}}
\def\hatq{\hat{q}}
\def\hatU{\hat{U}}

\def\hatA{\hat{A}}
\def\hatB{\hat{B}}
\def\hatC{\hat{C}}
\def\hatJ{\hat{J}}
\def\hatI{\hat{I}}
\def\hatP{\hat{P}}
\def\hatQ{\hat{Q}}
\def\hatU{\hat{U}}
\def\hatW{\hat{W}}
\def\hatX{\hat{X}}
\def\hatY{\hat{Y}}
\def\hatV{\hat{V}}
\def\hatt{\hat{t}}
\def\hatw{\hat{w}}

\def\hatp{\hat{p}}
\def\hatq{\hat{q}}
\def\hatU{\hat{U}}
\def\hatn{\hat{n}}

\def\hatphi{\hat{\phi}}
\def\hattheta{\hat{\theta}}

\def\iset{\mathcal{I}}
\def\fset{\mathcal{F}}
\def\pr{\partial}
\def\traj{\ell}
\def\eps{\epsilon}
\def\U{\hat{U}}

\def\U{U_{\rm cls}}
\def\P{P_{{\rm cls},\eta}}
\def\traj{\ell}
\def\cc{\cdot}

\def\DZ{D^{(0)}}
\def\Dcls{D_{\rm cls}}

\newcommand{\relmiddle}[1]{\mathrel{}\middle#1\mathrel{}}

\title{Quantum diffusion in the Harper model under polychromatic time-perturbation
}
\author{Hiroaki S. Yamada}
\affiliation{Yamada Physics Research Laboratory,
Aoyama 5-7-14-205, Niigata 950-2002, Japan}
\author{Kensuke S. Ikeda}
\affiliation{College of Science and Engineering, Ritsumeikan University, 
Noji-higashi 1-1-1, Kusatsu 525-8577, Japan}

\date{\today}
\begin{abstract}
Quantum dynamics of the Harper model with self-duality exhibits localized, 
diffusive, and ballistic states depending on the potential strength $V$.
By adding time-dependent harmonic perturbations composed of $M$ incommensurate frequencies, 
we show that all states of the Harper model transition to quantum diffusive states 
as the perturbation strength $\epsilon$ increases for $M \geq 3$.
The transition schemes and diffusion behaviors are discussed in detail and the phase diagram in the $(\epsilon,V)$ parameter space is presented.
\end{abstract}



\maketitle


\section{Introduction}
\label{sec:intro}
One-dimensional random lattice systems exhibit the Anderson localization. However, 
the application of  coherent periodic time-dependent perturbation may drastically alter 
the nature of localization \cite{yamada93}. Indeed, the time-dependent perturbation 
containing only a few incommensurate frequency components can destroy the 
localization. which has been investigated in detail 
\cite{yamada20,yamada21,yamada22,yamada23}.

On the other hand, it has been reported that a variety of  localized and 
delocalized behaviors  emerge in the wave-packet dynamics of one-dimensional  
quasi-periodic lattice systems, such as the Harper and kicked Harper models
\cite{artuso94,prosen01,kolovsky03,levi04,kolovsky12,wang13,qin14,cadez17,ravindranath21,
lakshminarayan03,mishra16,ray18,shimasaki22}.

However, it remains unclear whether quasi-periodic lattice systems exhibit a
transition to normal diffusion, similar to random lattice systems, under the
application of periodically oscillating coherent perturbations.  
The disorder of  quasi-periodic lattice system is much weaker than
in the random lattice systems, which is reflected in the fact that the former 
exhibits ballistic motion, in addition to localized behavior
depending on the strength of the on-site potential.

With this in mind, the purpose of this paper is to investigate the possibility 
of normal diffusion in the wave packet dynamics using the Harper model 
under smooth and periodically oscillating perturbations. 
(Hereafter referred  to as ``dynamical perturbations''.)
Although some studies have investigated the effect of the dynamical perturbation
 on quasi-periodic systems,
and emphasized the robustness of localized states 
\cite{Soffer03,Bourgain04,hatami16}, 
 whether such perturbations  can induce diffusive motion remains an open question.

Let the potential amplitude of the
quasi-periodic potential be $V$. Then the Harper model exhibits a remarkable feature.
There exist a critical value $V_c$, and, depending on $V$, the system takes three states, 
localized ($V>V_c$),  extended states ($V<V_c$), and critical state ($V=V_c$)
\cite{harper55,hofstadter76,aubry80,sokoloff85,evangelou00}.

In the context of solid state physics this  transition is interpreted
as an example of a metal-insulator transition (MIT).
Unlike the case of the 3D Anderson model with mobility edges, 
the transition in the Harper model arises from the self-duality of the model 
and is a sudden transition (abrupt transition) without the mobility edges.
(This self-duality is shown in Appendix \ref{app:duality}.)
A similar transition is also observed in the extended Harper model with self-duality
\cite{sun15,wang20,cai22,liu22, 
biddle09,molina14,rayanov15,ganeshan15,li17,major18,castro19,alexAn21}.
The above features are exactly reflected in the wavepacket dynamics.
\cite{hiramoto88,geisel91,wilkinson94,
ng07,larcher09,sarkar17}.
and the transition crossing over $V=V_c$ be regarded as a 
localization-ballistic transition (LBT).

We apply  the dynamical perturbation to the Harper model; the perturbation
composed of $M$ incommensurate frequency components with amplitude $\eps$,
 and elucidate how the three phases, i.e., the localized state(L), the ballistic state(B)
and the diffusive state(D) emerge in the two-parameter space $(\eps,V)$, by varying 
the number $M$ of frequencies, i.e., color number.  

When the oscillations are one-color ($M=1$) or two-color ($M=2$), 
the properties of the dynamics do not change significantly even when
$\eps$ is increased substantially, and 
only for $M\geq3$ do we observe that
the dynamics changes qualitatively and the quantum diffusion is induced. 
We emphasize that our system has neither spatial randomness nor
temporal non-analyticity such as kicked perturbation.
Only a few number of analytic, coherent, time-periodic perturbations can induce
an apparently time-irreversible diffusion phenomena.

More precisely, for $M\geq3$, 
both  localized state (which appears in the Harper model
for $V>V_c$) and the ballistic state (for $V<V_c$) undergo 
a transition to a quantum diffusive state.
We refer to the former as the localized-diffusion transition (LDT) 
and the latter as the ballistic-diffusion transition (BDT).
Particular attention is payed for the special critical case $V=V_c$, in which
a new type of transition may exist.
We compare the results with those observed in the KHM \cite{yamada23}.

The latter part of our paper is organized as follows:
In Sec.\ref{sec:model} we introduce the model 
and describe the numerical  methods to study the localization 
and delocalization phenomena dynamically.
In Sec.\ref{sec:Transition}, 
the original two phases of Harper model ,i.e, localization and ballistic motion,
undergoes transition to diffusive state by increasing the strength $\eps$ of
the dynamical perturbations.
Number of frequencies $M$ contained by the dynamical perturbation is crucial
for the presence of transition. As a special case the system with no static potential 
($V=0$) is also discussed.
In Sec.\ref{sect:typeAB} we discuss how the transition among the three phases, i.e.,
the localization phase, the diffusion phase  and the ballistic phase occurs by 
scanning the strength $V$ of static potential. Based upon such observations
we present the phase diagram of the three phases.
In Sec.\ref{sec:diffusion} the nature of diffusive property realized after the
transition is discussed in detail. The diffusion constant, which first increases
with the perturbation strength finally turns to decrease.
Finally, in Sec.\ref{sec:critical-region}, we show a complicated behavior 
observed for the critical region, i.e.,the boundary region between the 
localization phase and the ballistic phase. Transition between two different 
types of diffusion is suggested. 
A summary is given in the last section.

\section{Model}
\label{sec:model}
We consider the Harper model described by the following
Hamiltonian with the dynamical perturbation:
\beq
H(t)&=&\sum\limits_{n=1}^N {|n \rangle v(n)[V+\eps f(t)] \langle n|} \nn \\
 &+& T\sum\limits_{n}^N ({|n \rangle \langle n+1|+
|n+1 \rangle \langle n|}).
\label{eq:Hamiltonian}
\eeq
The on-site energy sequence is 
\beq
  v(n)=2 \cos(2\pi Q n+\theta), 
\eeq
where $\{|n \rangle \}$ is an orthonormalized basis set and the $Q$ is an irrational number. 
$V$ is potential strength, and $T$ denotes the hopping energy 
between adjacent sites, respectively.
We tale $Q=\frac{\sqrt{5}-1}{2}$, and $T=-1$ throughout the present paper.
Although $\theta$ is an arbitrary phase of the potential, 
and it is used for an average over it.
$\eps$ is the strength of the dynamical perturbation whose functional form
is given by a sum of the incommensurate harmonic oscillations,  
 \beq
 \label{eq:perturbation}
f(t)=\frac{1}{\sqrt{M}} \sum_i^M\cos(\omega_i t+\varphi_i), 
\eeq
where $M$ is the number of frequency components and 
the frequencies $\{ \omega_i\}(i=1,...,M)$ are taken as mutually incommensurate 
numbers of order $O(1)$.
Note that the long-time average of the total power of the perturbation is normalized to 
$\overline{f(t)^2}=1/2$ and $\{\varphi_i \}$ are the initial phases.
For  the long-time behavior, the choise of initial phase $\varphi_i$ is irrelevant, and so
we generally take $\varphi_i=0$, but we take $\varphi_i$ as random values if necessary.

For the unperturbed case ($\eps=0$),  this model was introduced as an model of electron 
in a two-dimensional crystal under a strong
external  magnetic field.
[Note that there are also references that describes this model as Aubry-Andre model (AA model) or Aubry-Andre-Harper model (AAH model). ]
Throughout this paper we take the relative strength $\eps/V$ instead of
$\eps$ itself if $V\ne0$.
We remark that, in our previous publications \cite{yamada99,yamada22}, $\eps$ 
was taken as the parameter characterizing the relative perturbation strength,
which corresponds to $\eps/V$ in the present paper.
The case of $V=0$  is very specific in the sense that no
site energy exist, and an ideal ballistic motion appears
if $\eps=0$. We are much interested in the effect of the dynamical 
perturbations on ideally ballistic motion, and we also discuss 
this case as a special case.

We set the initial wave packet $<n|\Psi(t=0)>=\delta_{n,n_0}$
 localized at a single site $n_0$, 
and calculate the time evolution of the wavefunction $|\Psi(t)>$ 
using  the Schrodinger equation: 
\beq
i\hbar\frac{\pr |\Psi(t)>}{\pr t}=H(t)|\Psi(t)>.
\eeq
We monitor the spread of the wavefunction in the site space by the 
mean square displacement (MSD),
\beq
m_2(t) = \sum_{n}(n-n_0)^2 \left< |\phi(n,t)|^2 \right>, 
\eeq
where $\phi(n,t)=<n|\Psi(t)>$ is the site representation of the wave function.
Numerical calculations were performed using second-order symplectic integrator 
with stable time increments $\Delta t=0.005 \sim 0.02$.
We mainly use the system size $N=2^{13}-2^{17}$, 
and $\hbar=1/8$.

For the localized, ballistic, and diffusive motion, $m_2(t)$ changes as
$m_2(t) \sim t^0, t^2$, and $t^1$, respectively. In addition, 
an anomalous diffusion 
\beq
 m_2(t) \sim t^\alpha 
\eeq
characterized by the non-integer diffusion index $\alpha$ may appear especially
at the critical point of the localization-diffusion transition (LDT) and 
the ballistic-diffusion transition (BDT).

We can extend $\alpha$ as a time-dependent index, which characterizes
the tangent in the double logarithmic-plots of $m_2$ vs $t$.
The time-dependent diffusion index $\alpha$ 
is numerically calculated as 
\beq
 \alpha(t)=\frac{d\log\ovl{m_2(t)}}{d\log t}
\eeq
by using the locally time-averaged MSD $\ovl{m_2(t)}$
which is taken over characteristic time scales.
This quantity is useful for describing the overall appearance 
of the transitions to the diffusion state.
In the case of the LDT, for example, the index $\alpha(t)$ decreases toward
$0$ indicating the localization if $\eps$ is small enough, whereas
it increases toward $1$ implying the normal diffusion, and there
may be certain $\eps_c$ at which $\alpha(t)$ tends to a finite constant value
$0<\alpha(t)=\alpha_c<1$, as were confirmed  in our previous publications 
\cite{yamada20,yamada22}. 
We expect similar behavior in the index $\alpha(t)$ for the case of the BDT.

\section{Transition to diffusion due to increse in $\eps$}
\label{sec:Transition}

\subsection{Localization side: $V > V_c=1$}
In this subsection, we set the potential strength $V$ to a localized state and 
examine the dynamics of the wave packet due to the dynamical perturbation.

\subsubsection{Absence of transition: $M=1,2$}
\label{sec:Localization}
Before proceeding to the perturbation-induced transition 
observed for $M\geq3$, which is one of the main topic of this paper, 
let us summarize the localization phenomena for $M=1$ and $M=2$ in this section.

\begin{figure}[htbp]
\begin{center}
\includegraphics[width=4.4cm]{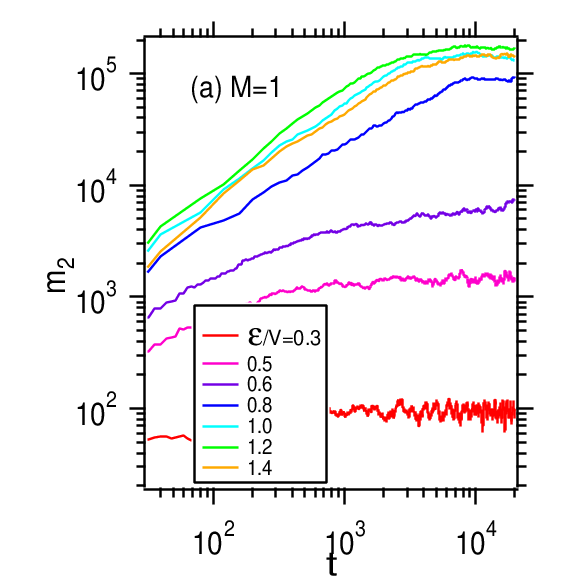}
\hspace{-5mm}
\includegraphics[width=4.4cm]{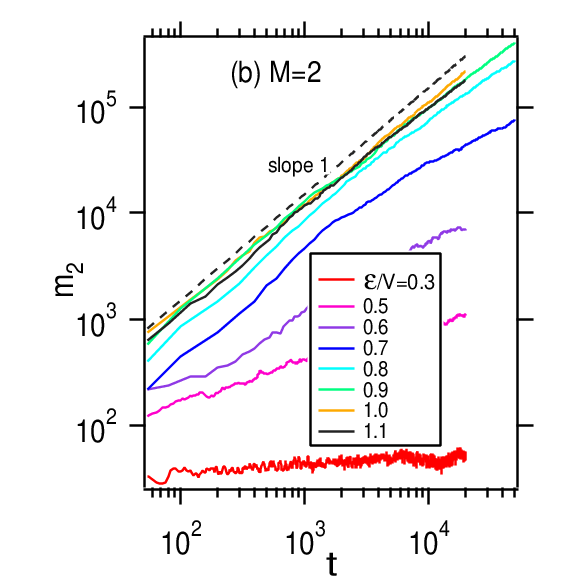}
\caption{(Color online)
\label{fig:Hc1-1}
The double logarithmic plots of $m_2$ as a function of $t$ 
for several values of the perturbation strength $\eps$
in a case of the potential strength $V=1.3$.
(a)$M=1$ and (b)$M=2$. 
Sample averages were made for 10 potential phases $\theta$.
$\hbar=1/8$. The subsequent numerical results are also processed
in the same way.
}
\end{center}
\end{figure}

In the cases, $M=1$ and $M=2$, with a potential strength fixed at $V=1.3$
the MSD $m_2(t)$ for  different $\eps$
are shown in Fig.\ref{fig:Hc1-1}(a) and (b), respectively.
Localization is maintained at least when the relatively small perturbations 
is applied.
And then, as the perturbation strength is further increased, 
it remains localized for $M=1$. 
On the other hand, 
for $M=2$, the dynamics asymptotically 
approaches normal diffusive behavior, $m_2 \propto t^1$, as $\eps$ grows,
as can be seen from Fig.\ref{fig:Hc1-1}(b).

\begin{figure}[htbp]
\begin{center}
\includegraphics[width=6.5cm]{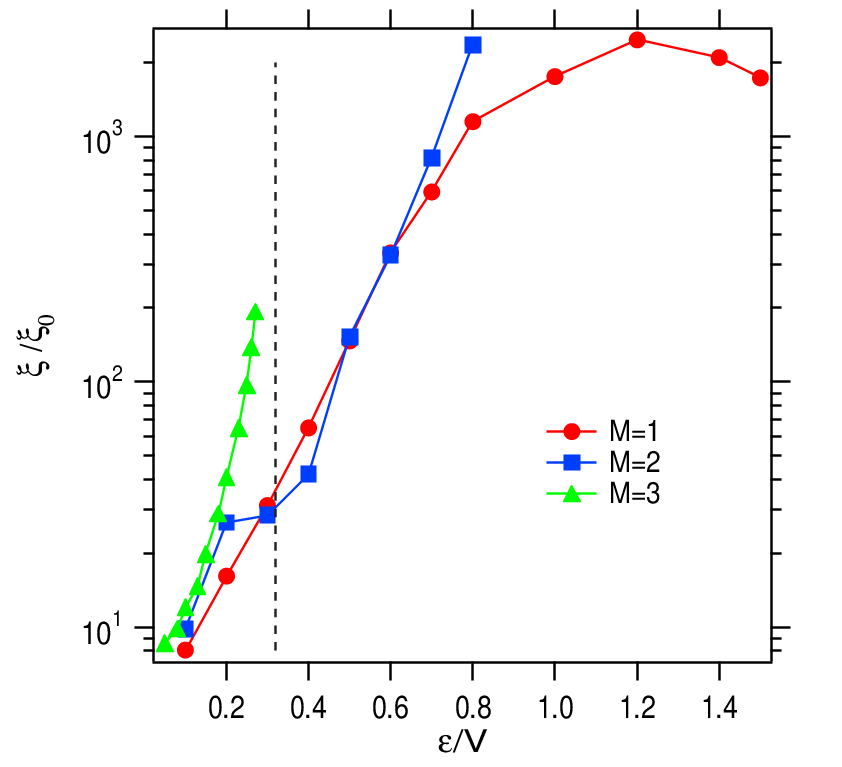}
\caption{(Color online)
\label{fig:loclen-c1}
The scaled dynamical localization length $\xi/\xi_0$ as a function of $\eps/V$
 for $M=1,2,3$ in a case of $V=1.3$. 
The dotted line represents critical strength $\eps/V=0.32$ for $M=3$.
Note that the vertical axis is logarithmic scale.}
\end{center}
\end{figure}

The $\eps-$dependence of the numerically computed 
dynamical localization length $\xi=\sqrt{m_2(t\to \infty)}$ 
in the cases, $M=1$ and $M=2$, 
are shown in Fig.\ref{fig:loclen-c1}.
The localization length $\xi$ scaled by the  localization 
length $\xi_0(=1/\ln |V|)$ of the unperturbed Harper model is shown.
In both cases, for $\eps/V \lesssim 0.8$, it increases exponentially.
\beq
 \xi(V,\eps) \simeq \xi_0(V) \e^{c\eps}, 
\eeq
where $c$ is a constant.\\

In the case of $M=1$, 
the localization length 
reaches a maximum at $\eps/V \simeq 1.2$ 
and then begins to decrease.
On the other hand, in the case of $M=2$, 
the localization length becomes so large that it cannot be captured
numerically.
Such $\eps-$dependency of the localization length are similar to 
that found in the Anderson model and the Anderson map 
model \cite{yamada20,yamada22}.

See Appendix \ref{app:duality} for the self-duality of the Harper model 
and the localization length $\xi_0$ that can be derived from it.

\subsubsection{Localization-diffusion transition (LDT): $M \ge 3$ }
\label{sec:LDTBDT}
Here, we examine the LDTs and BDTs 
that emerge in $M\ge3$ due to the changes in $\eps/V$ and $V=0$, 
and finally give an outline of the phase diagram in $(\eps,V)$ space.

In this subsection, we fix the potential parameter $V$ 
to some values in the  localized region $V>V_c=1$ of the Harper model, 
and investigate how dynamics transition 
 to the diffusive state by changing the parameters 
$M(\ge 3)$ and $\eps$ of the polychromatic perturbation.\\

\begin{figure}[htbp]
\begin{center}
\includegraphics[width=4.4cm]{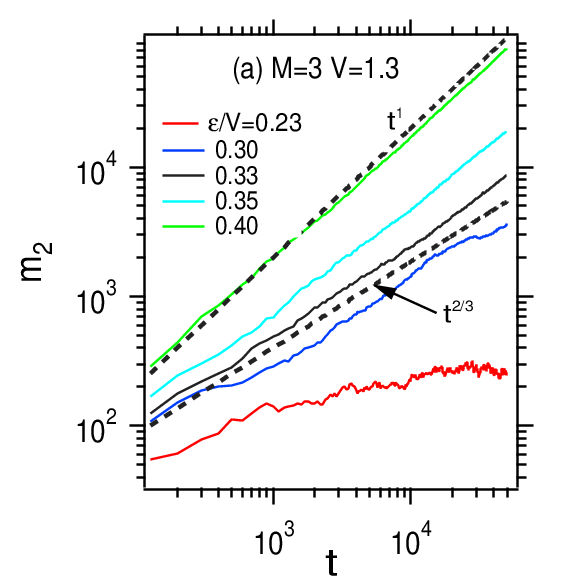}
\hspace{-5mm}
\includegraphics[width=4.4cm]{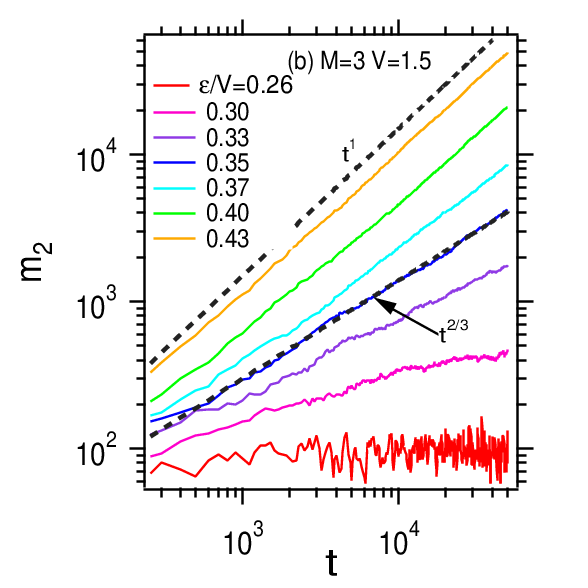}
\hspace{+5mm}
\includegraphics[width=4.4cm]{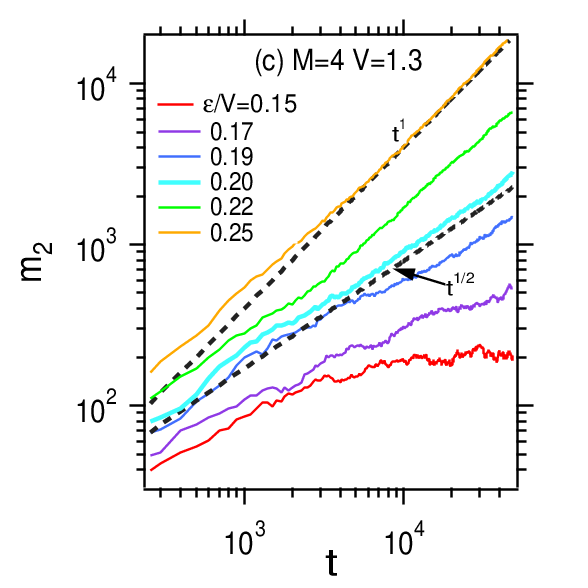}
\hspace{-5mm}
\includegraphics[width=4.4cm]{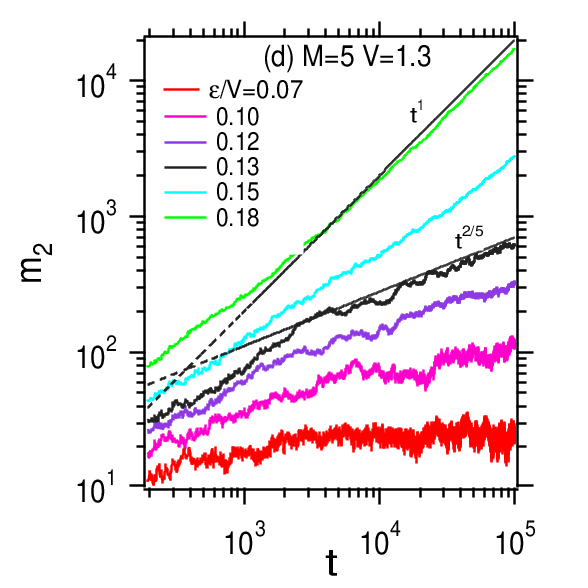}
\caption{(Color online)\label{fig:c345}
The double logarithmic plots of $m_2$ as a function of $t$ 
for several values of $\eps$ and $V$.
 (a)$M=3$,$V=1.3$ (b)$M=3$,$V=1.5$ (c)$M=4$,$V=1.3$ (d)$M=5$, $V=1.3$. 
The dashed lines indicate $m_2 \propto t^1$ and $m_2 \propto t^{2/M}$ in each case.
}
\end{center}
\end{figure}

Figure \ref{fig:c345} shows the time evolution of $m_2(t)$ with increasing the perturbation 
strength $\eps$ for $M=3,4,5$.
It can be seen from Fig.\ref{fig:c345}(a) and (b) that
in the case of $M=3$,  for both $V=1.3$ 
and $V=1.5$,  subdiffusion of $m_2 \sim t^{\alpha}, \alpha \simeq2/3$, 
is realized at  $\eps/V\simeq 0.32 (\eps_c\simeq 0.4)$ and 
$\eps/V\simeq 0.35 (\eps_c\simeq 0.52)$, respectively.
On the other hand, for $\eps>\eps_c$, $m_2(t)$ becomes
showing the normal diffusion $m_2 \sim t^1$, and 
for $\eps<\eps_c$, it tends to localized for a long time.
The localization lengths for $\eps>\eps_c$ are plotted in the Fig. \ref{fig:loclen-c1}.

Furthermore, a similar transition is observed for larger values of $M$: 
as seen in Fig.\ref{fig:c345}(c) and (d). the critical subfiffusion 
occurs for $M=4$ and $M=5$ with exponents  
$\alpha \simeq 2/4$ and $\alpha \simeq 2/5$, respectively. 

\begin{figure}[htbp]
\begin{center}
\includegraphics[width=4.4cm]{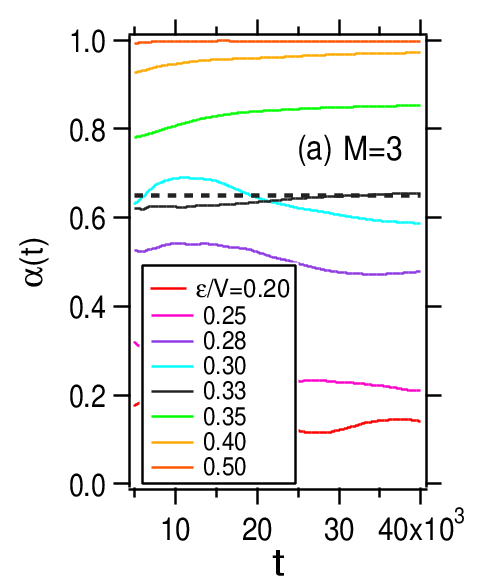}
\hspace{-5mm}
\includegraphics[width=4.4cm]{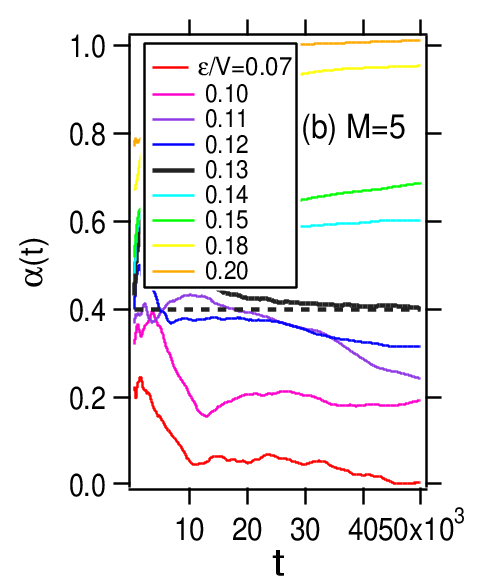}
\caption{(Color online) \label{fig:A-alphar-c3c5}
The time-dependence of $\alpha(t)$ 
for various strength of $\eps$ in the case of $V=1.3$. 
(a)$M=3$ and  (b)$M=5$.
The dotted lines indicate $\alpha(t)=2/M$.
}
\end{center}
\end{figure}

To confirm the above observations, the time variation of the instantaneous 
diffusion index $\alpha(t)$ is shown in Fig.\ref{fig:A-alphar-c3c5}. 
With an increase in $\eps$, 
it changes from $\alpha(t)\to0$
to $\alpha(t)\to1$ for $t\to\infty$, supporting strongly the presence of critical 
subdiffusion in which $\alpha(t)$ keeps a constant fractional value.
The $M$-dependence of the critical value $\eps_c$ at the transition point is 
approximately monotonically decreasing for $M$: 
\beq
    \eps_c \sim \frac{1}{M-2} (M\geq3), 
\eeq
which is similar to the case of Anderson model \cite{yamada21}.

Thus, it is suggested that the LDT is a universal phenomenon caused 
by the polychromatic time-perturbation to the localized state.
In fact, similar LDTs are seen in the localized state of quasi-periodic system
 that differ from the Harper model, 
as shown in Appendix \ref{app:maryland}.

\subsection{Ballistic side: $0<V<V_c=1 $ }
In this subsection, we set the potential strength $V$ to the ballistic state and 
examine the dynamics of the wave packet due to the dynamical perturbation.

\subsubsection{Absence of transition: $M=1,2$}

Although the time-periodic perturbation destroys the self-duality
of the original Harper equation, there should be a correspondence
between the localized region $V>1$ and the ballistic region $V<1$.
We expect that the ballistic motion is maintained for $M=1$ and $M=2$.
Indeed, for $M=1$, the ballistic spreading 
is not broken although  $\eps$ is increased large enough. 
The case of $M=2$ is critical and asymptotically approaches $\alpha \to 1$ 
with increase in $\eps$, but the ballistic spreading is not broken although  
$\eps$ is increased large enough.

\subsubsection{Ballistic-diffusion transition (BDT): $M\ge 3$}

For $M\ge 3$ we may expect the existence of the ballistic-diffusion transition (BDT) 
for $0<V<1$ corresponding to the occurrence of the LDT for the $V>1$ side.

Indeed, in the kicked Harper model (KHM), 
a time-discrete version of the Harper model, 
existence of  BDT and LDT 
by the time-periodic perturbation was confirmed \cite{yamada23}.

\begin{figure}[htbp]
\begin{center}
\includegraphics[width=4.4cm]{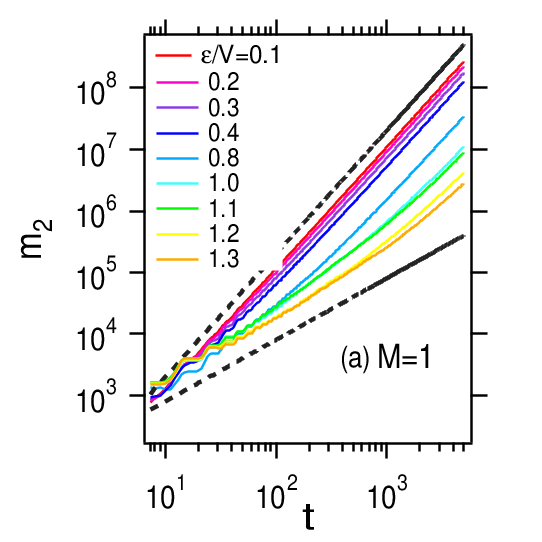}
\hspace{-5mm}
\includegraphics[width=4.4cm]{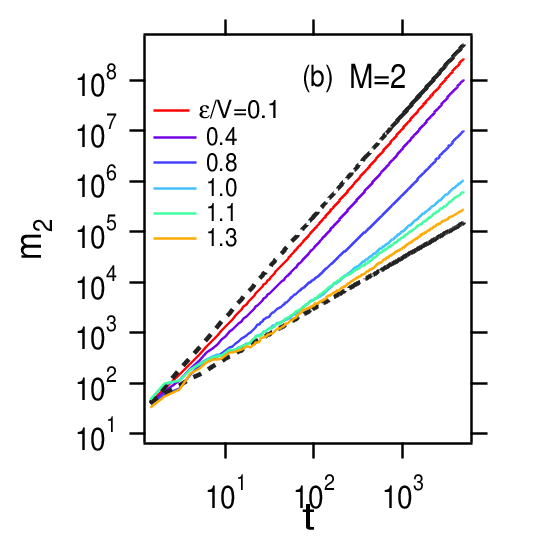}
\hspace{5mm}
\includegraphics[width=4.4cm]{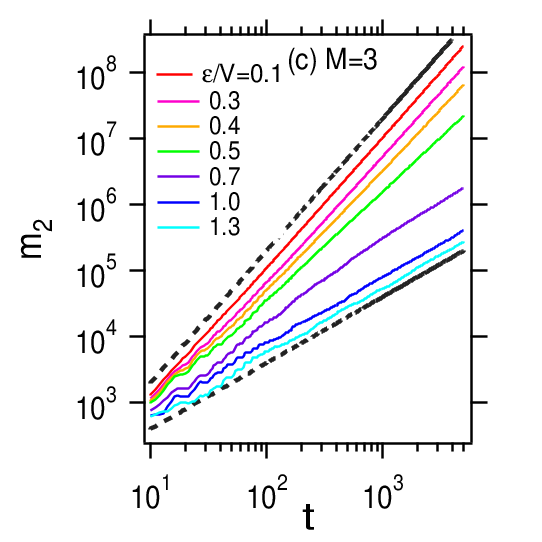}
\hspace{-5mm}
\includegraphics[width=4.4cm]{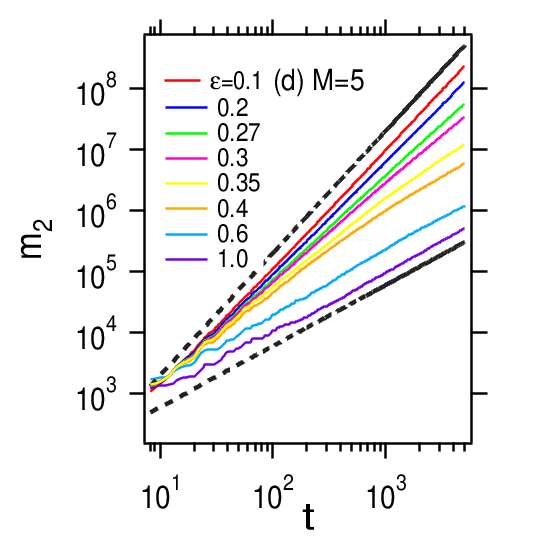}
\caption{(Color online) \label{fig:c1c2c3c5-BDT}
The double-logarithmic plots of $m_2(t)$ as a function of $t$ 
at various strength of $\eps$
 in the case of $V=0.7$.
 (a)$M=1$, (b)$M=2$, (c)$M=3$ and  (d)$M=5$.
The solid black lines indicate normal diffusion $m_2 \propto t^1$ 
and ballistic spreading $m_2 \propto t^2$.
Note that the axes are logarithmic.
}
\end{center}
\end{figure}

\begin{figure}[htbp]
\begin{center}
\includegraphics[width=4.4cm]{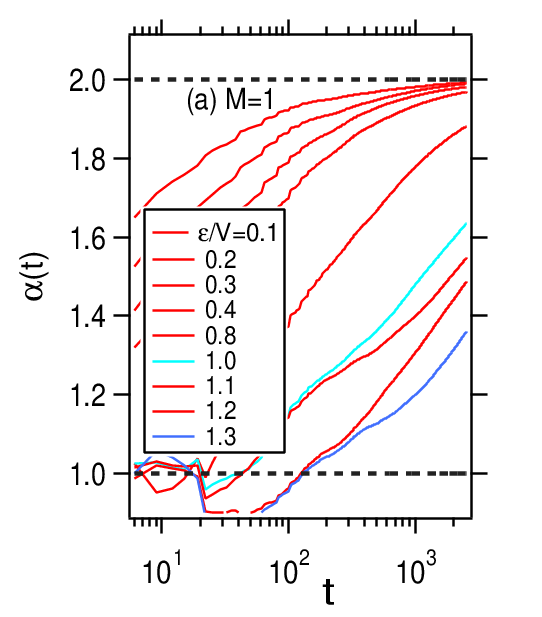}
\hspace{-5mm}
\includegraphics[width=4.4cm]{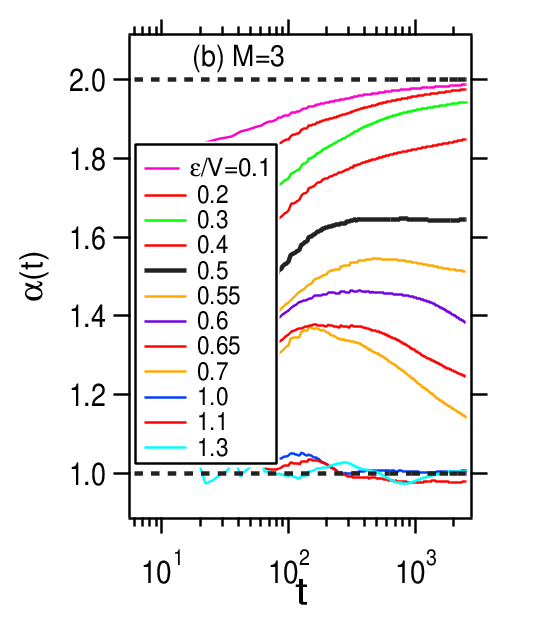}
\caption{(Color online)\label{fig:v07-alpha}
The time-dependence of $\alpha(t)$ 
at various values of $\eps$ in the case of $V=0.7$.
(a)$M=1$, (b)$M=3$. 
The dashed lines indicate normal diffusion $\alpha(t)=1$ 
and ballistic spreading $\alpha(t)=2$.
}
\end{center}
\end{figure}

Figure \ref{fig:c1c2c3c5-BDT} shows the time evolution of MSD
as the time-periodic perturbation is applied to the ballistic state
of the Harper model for $V \sim 0.7$. The double log plot shows
that the increasing power of  MSD changes from 2 to 1 with an
increase in $\eps$. To confirm this, the time variation 
of the diffusion index $\alpha(t)$ is displayed in Fig.\ref{fig:v07-alpha}
for several values of $\eps$. Evidently, there exist a critical value 
$\eps/V=\eps_b/V$ at which $\alpha(t)$ keeps a constant fractional value
between 1 and 2. Above and below $\eps_c/V$, 
$m_2(t)$ asymptotically approaches toward $1$ or $2$, respectively.

The case $V=0$ is particular in the sense that there is no scattering
potential and the particle is completely free and show an ideal  ballistic motion
if $\eps=0$. 
It will be interesting to see whether the periodic perturbation induces normal 
diffusion in such a case.
The last subsection describes this specific case.

\subsection{Specific case: $V=0$} 
\label{sec:B-type}

In the limit $V \to \infty$ the critical point $\eps_c$ of LDT tends to diverge. Correspondingly,
the critical point $\eps_b$ of BDT may diverge and there may be no transition in the
limit of $V\to 0$, which means that the periodic perturbations composed of
few frequencies can not make a completely free particle diffusive.

\begin{figure}[htbp]
\begin{center}
\includegraphics[width=4.4cm]{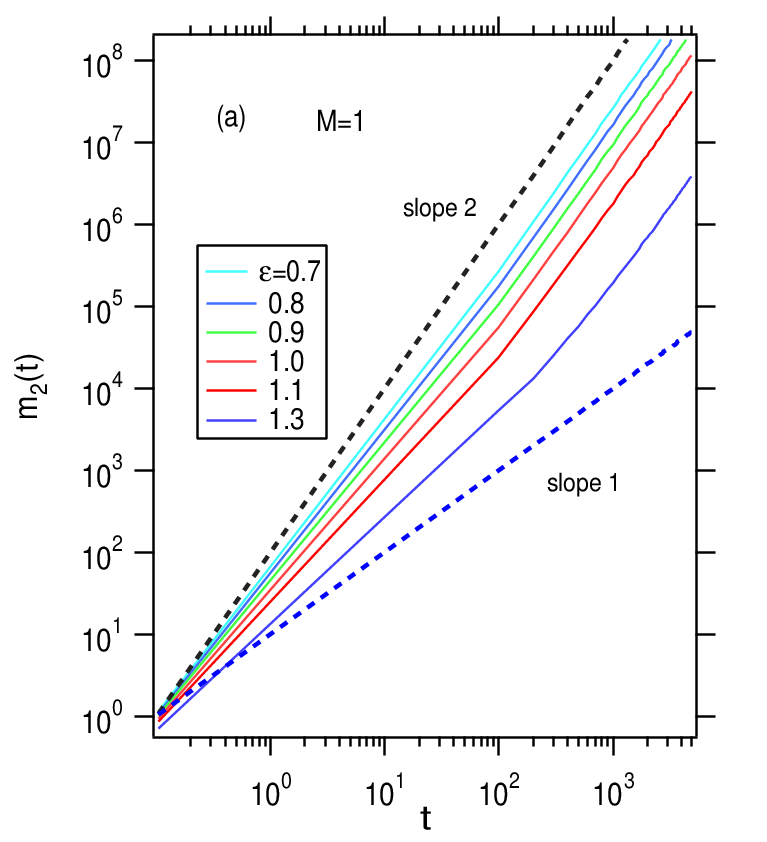}
\hspace{-5mm}
\includegraphics[width=4.4cm]{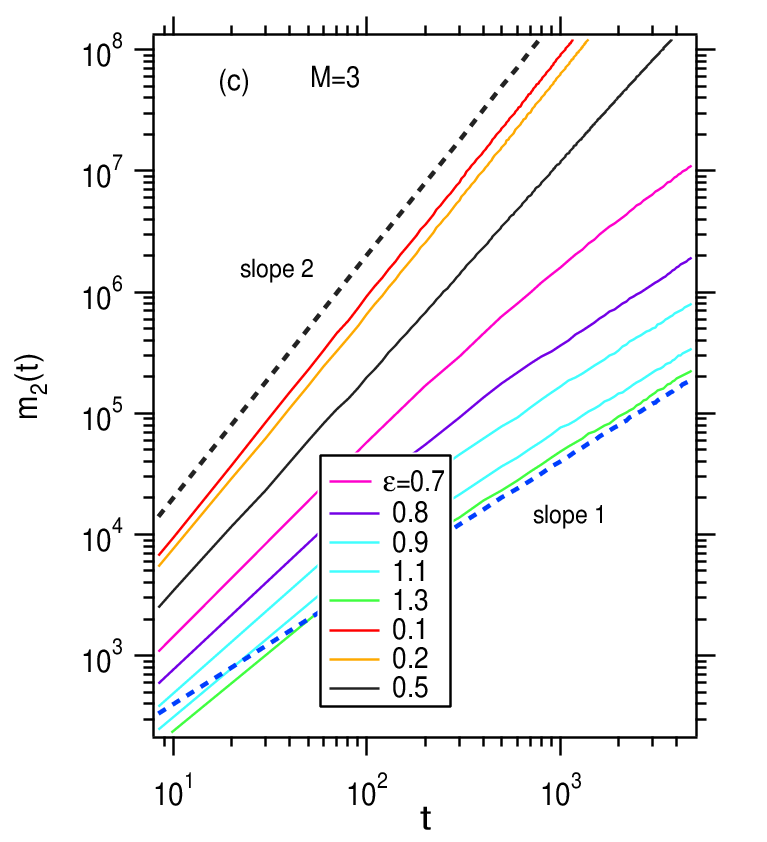}
\hspace{5mm}
\includegraphics[width=4.4cm]{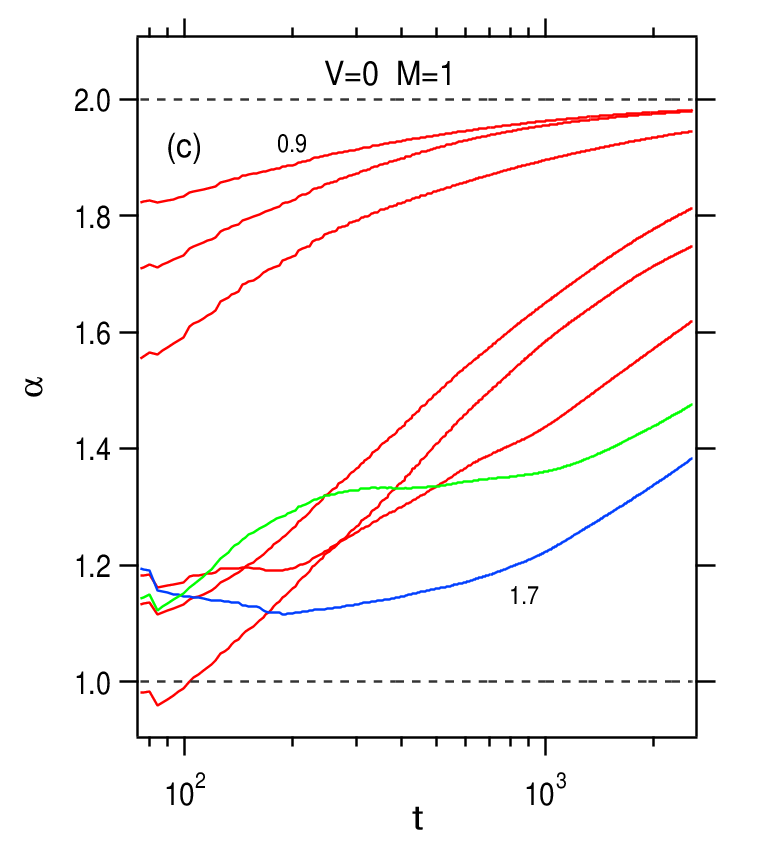}
\hspace{-5mm}
\includegraphics[width=4.4cm]{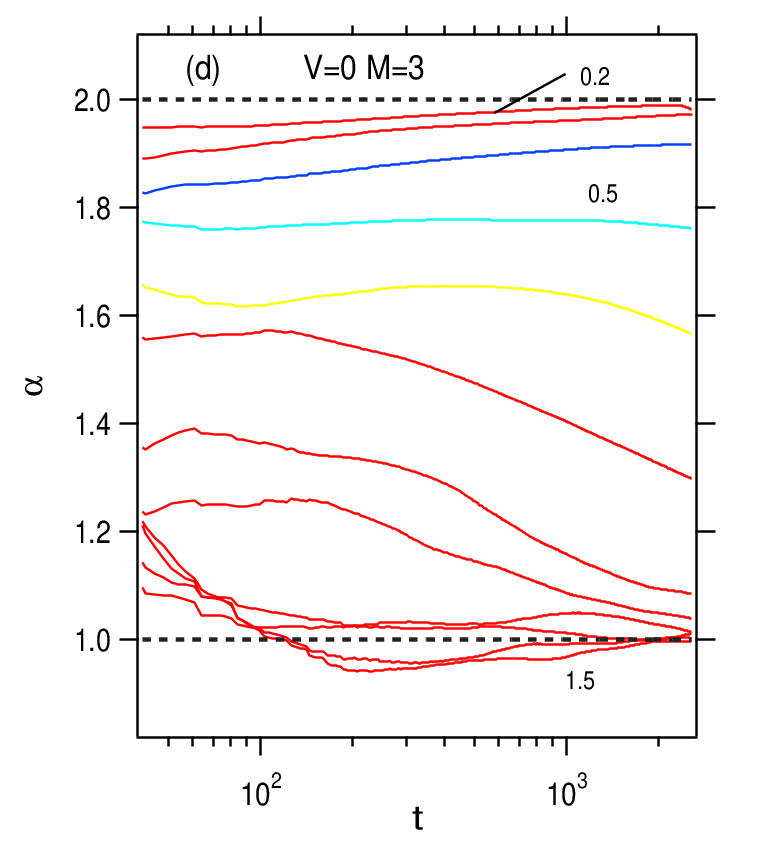}
\caption{(Color online)
\label{fig:B-tcont-c1c2}
The double-logarithmic plots of $m_2$ as a function of $t$ 
for various strength $\eps=0.7,0.8,0.9,1.1,1.3$ from top to bottom, 
 in the case of  $V=0$. 
 (a)$M=1$, (b)$M=3$. $\hbar=1/8$.
The dashed lines indicate normal diffusion $m_2 \sim t^1$ 
and ballistic spreading $m_2 \sim t^2$.
The time-dependence of $\alpha(t)$  at various values of $\eps$ 
in the case of $V=0$ and  (a)$M=1$, (b)$M=3$, where 
$\eps$ is changed from $\eps=0.9$ to $1.7$ in the panel (c),
and from $\eps=0.2$ to $1.5$ in the panel (d), respectively.
The dotted lines indicate $\alpha=1$ and $\alpha=2$.
}
\end{center}
\end{figure}

For $M=1$ the ballistic motion is not destroyed by applying the
time-periodic perturbation and follows the feature discussed in the previous subsection.
Figure \ref{fig:B-tcont-c1c2}(a),(c) depict the temporal evolution of MSD and $\alpha(t)$ 
at various values of $\eps$ for $M=1$.  The index $\alpha(t)$ is also always goes
toward the ballistic $\alpha=2$.

However, for $M\geq3$,  the transition from ideal ballistic 
motion $m_2\propto t^2$ to diffusion $m_2\propto t^1$ (BDT) occurs even though 
$V=0$, as is manifested in \ref{fig:B-tcont-c1c2}(b),(d), 
a critical anomalous diffusion of the exponent
$\alpha(t)=\alpha_c\sim 1.75$  is seen at $\eps=\eps_b\sim 0.5$.
This fact implies that excitation of only a few coherent phonon mode 
is enough to convert the ballistic motion of electron to irreversible diffusive motion. 

\section{The phase diagram: succession of transitions among three states by varying $V$}
\label{sect:typeAB}

If $M\geq 3$, there are three phases  i.e., the localized (L), the ballistic (B),
and the diffusive (D) states. To investigate the relative arrangement of the 
three phases, we first change the parameter $V$ fixing the dynamical 
perturbation strength $\eps$, and observe what occurs.

\begin{figure}[htbp]
\begin{center}
\includegraphics[width=6.5cm]{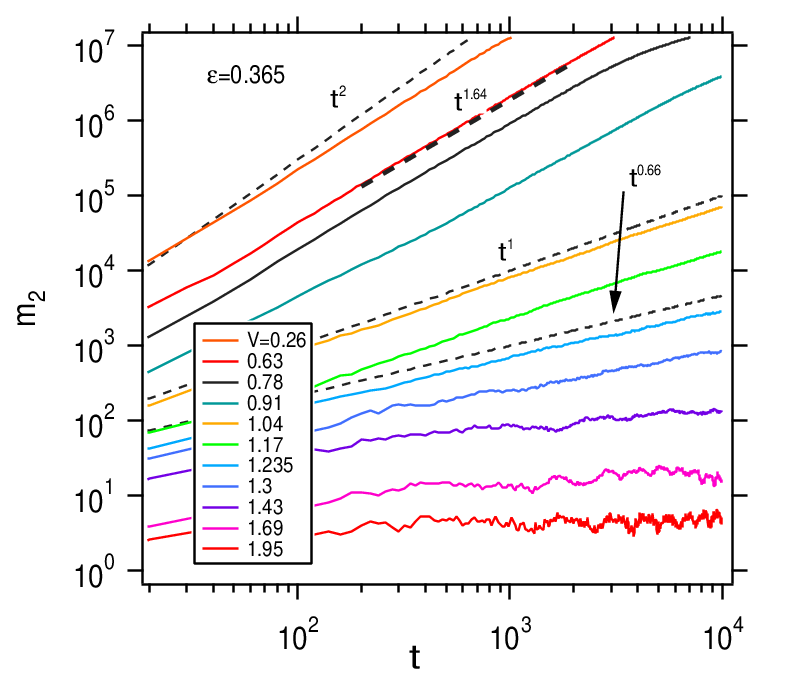}
\caption{(Color online)
\label{fig2:eps-L-LDT}
The double logarithmic plots of $m_2$ as a function of $t$ 
for some values of the parameter $V$
 in the case of $M=3$ and $\eps=0.365$.
The solid lines have slope 0.66, 1.0, 1.64  and 2, respectively.
}
\end{center}
\end{figure}

Figures \ref{fig2:eps-L-LDT} is the result for $\eps=0.365$, 
starting from phase B close to $V=0$ and increasing $V$, 
we first encounter with a transition from phase B ($m_2 \sim t^2$) 
to phase D ($m_2 \sim t^1$) via the critical anomalous diffusion 
$m_2 \sim t^{1.64}$ occurs. With a further increase 
$V$, the second transition via the critical anomalous diffusion 
$m_2 \sim t^{0.66}$ is observed, and finally the localized phase appears.
Such a feature do not change if we start the phase B at $V=0$
and increase $V$.
Such a feature do not change if we vary $\eps$ in the range below the 
critical value $\eps_b\sim 0.5$ at $V=0$, as discussed in section \ref{sec:B-type}

\begin{figure}[htbp]
\begin{center}
\includegraphics[width=7.0cm]{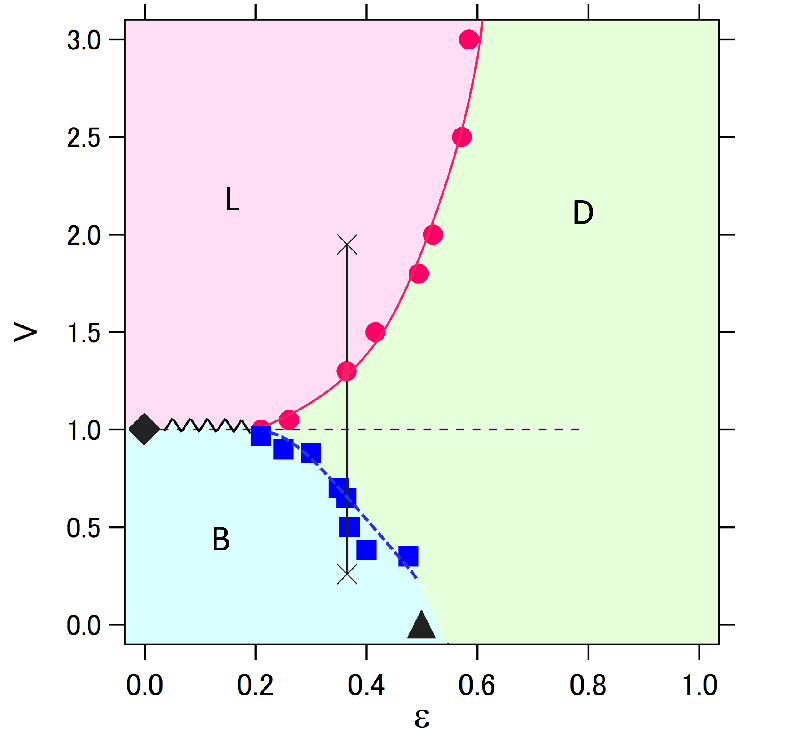}
\caption{(Color online)\label{fig:image-fig3}
The schematic phase diagram in the $(\eps,V)$ plane  
for the perturbed Harper model with $M=3$.
The wave packets are localized in the phase $L$,  are diffusive in the phase $D$,
and are ballistic in the phase $B$.
Some numerical results of $\eps_c$ and $\eps_b$ are plotted in the diagram 
by some symbols.
When $V$ is varied along the line connecting the two cross-marked points in the
L phase and B phase, the BDT and LDT occurs successively. 
as  shown in Fig.\ref{fig2:eps-L-LDT}.
The area indicated by the jagged lines is a complex area 
where phases L, B, and D are mixed.
The filled diamond and triangle indicate MIT point for $V=V_c=1$
and BDT point for $V=0$, respectively. 
}
\end{center}
\end{figure}

From the above observations one can imagine that the arrangement
of the three phases in the $(\eps,V)$ space is shematically as shown
in Fig.\ref{fig:image-fig3}.
Note the two critical curves, namely, $\eps_c$ curve of LDT and the 
$\eps_b$ curve of BDT are explicitly displayed in the Fig.\ref{fig:image-fig3}.
The three states, i.e., localized, diffusive, and ballistic, are denoted 
by L, B and D respectively, and they are color-coded.
As a result, in the Harper model of $M\geq3$ any state of the 
wave packet propagation is led to the quantum normal diffusion with 
increase in the perturbation strength $\eps$.
It can be seen that as $M$ increases, the areas of 
phase L and B tend to shrink and the area of phase 
D tends to expand.

As is shown in Fig.\ref{fig:image-fig3}, if $V$ is increased with 
fixed $\eps$ along the line connecting the
two cross marked points of B and L, the BLT or LBT is realized
successively if the $\eps$ is appropriate.
%
However, as $\eps$ gets smaller, the two critical curves, i.e., the $\eps_c$ curve 
and the $\eps_b$ curve come together, and the three phases L,D,and B 
are mixed on the line $V=1$, and very complex dynamical behaviors may be observed
along the jugged line in Fig.\ref{fig:image-fig3} which indicates
the region of the two critical curves coming close together.
This will be discussed in section  \ref{sec:V=Vc}.


\section{Diffusive phase}
\label{sec:diffusion}
In this section, we summarize the diffusion properties observed for
$\eps>\eps_c$, including the critical case $V=V_c$. 
This diffusion coefficient $D$ used in this section is determined by
\beq
\label{eq:Diffusion}
 D=\lim_{t \to \infty}\frac{m_2}{t}.
\eeq
from numerical results.

\subsection{$\eps-$dependence of the diffusive behavior} 

\begin{figure}[htbp]
\begin{center}
\includegraphics[width=6.5cm]{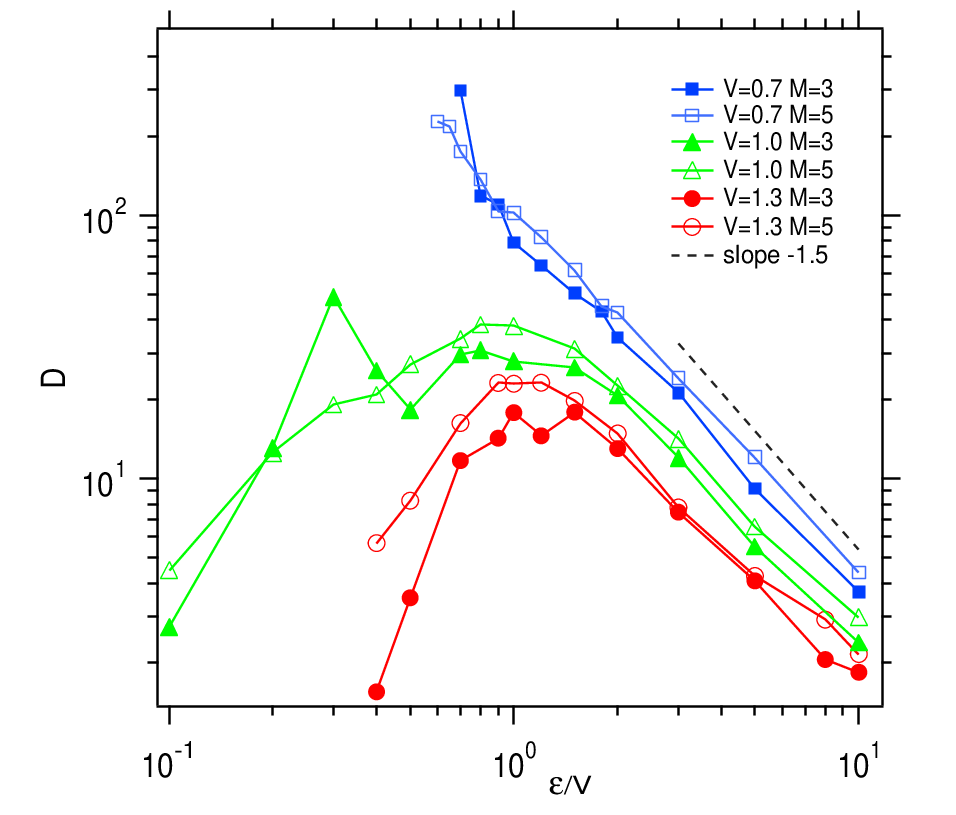}
\caption{\label{fig:unified-D}(Color online)
Diffusion coefficient $D$ as a function of $\eps/V$ 
for three representative values of $V$ i.e., $V=0.7(<V_c)$, $V=1.0(=V_c)$ 
and $V=1.3(>V_c)$, which are colored by blue, green and red, respectively,
where $M=3$ and $M=5$.  The broken line indicates slope $-3/2$.
 Note that the both axes are in logarithmic scale.
}
\end{center}
\end{figure}

 Figure \ref{fig:unified-D} show the diffusion coefficients as a function 
of the perturbation strength $\eps(>\eps_c)$ for $M=3$ and $M=5$ 
at typical values of $V$ in the region $V>V_c(=1)$, $V<V_c$ and at the 
critical value $V=V_c$.
The blue, green, and red respectively corresponds to the three regions  
$V>V_c$, $V<V_c$ and $V=V_c$.

We first discuss the localization regime $V>V_c$, where the unperturbed limit ($\eps=0$) 
exhibits localized state of Harper model.
A remarkable character of $D$ in this regime is that $D$ first increases 
from 0 $\eps$ exceeds $\eps_c$. As $\eps$ goes over a characteristic value $\eps^*$,
$D$ decreases as is demonstrated by red circles in Fig.\ref{fig:unified-D}. The
maximum appears at the relative strength $\eps/V=\eps^*/V(\simeq 1.2)$.
$D$ decreases following a power law $D \propto \eps^{-1.5} $ beyond 
$\eps^*$. Such a power decrease was observed also in random system Ref.\cite{yamada21}.

Next,  we discuss the ballistic regime $V<V_c=1$. As is shown by red circles, the diffusion 
coefficient behaves quite simply, if $V$ is not close to the critical value 1.  
It decays monotonically with increase in $\eps(>\eps_b)$, following the power 
law $D \sim \eps^{-3/2}$ if $\eps$ increase large enough. 
This decrease is the same as in the case of $V>V_c$ mentioned above.
In the ideal free particle limit $V=0$, the above features 
almost hold. See Appendix \ref{app:DofV0} for the 
$\eps-$dependence of the Diffusion coefficient $D$.

Finally, the critical value $V=V_c$ is a particular case in which the 
diffusion exists even at $\eps=0$ due to the self-duality of the Harper model.
Roughly speaking, $D$ follows the case of $V>V_c=1$ as shown by 
green triangles: It increases from a finite value and decreases as 
$D \propto \eps^{-1.5}$.after $\eps$ exceeds a certain value.
See Appendix \ref{app:DofV0} for the MSD $m_2(t)$ in the case.
 
A closer observation, however, reveals that this region exhibits some 
anomalous features, as discussed in the Sect.\ref{sec:V=Vc}.

\subsection{$V-$dependence of the diffusive behavior}
\begin{figure}[htbp]
\begin{center}
\includegraphics[width=4.4cm]{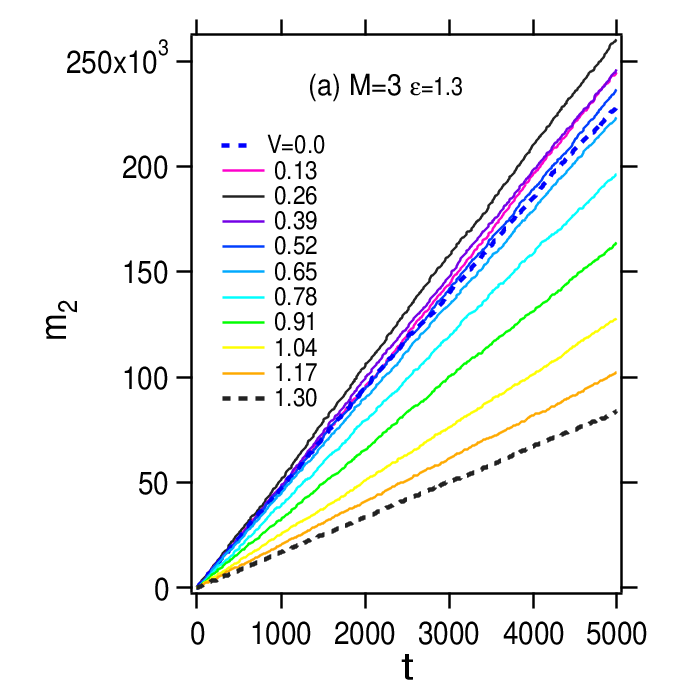}
\hspace{-5mm}
\includegraphics[width=4.1cm]{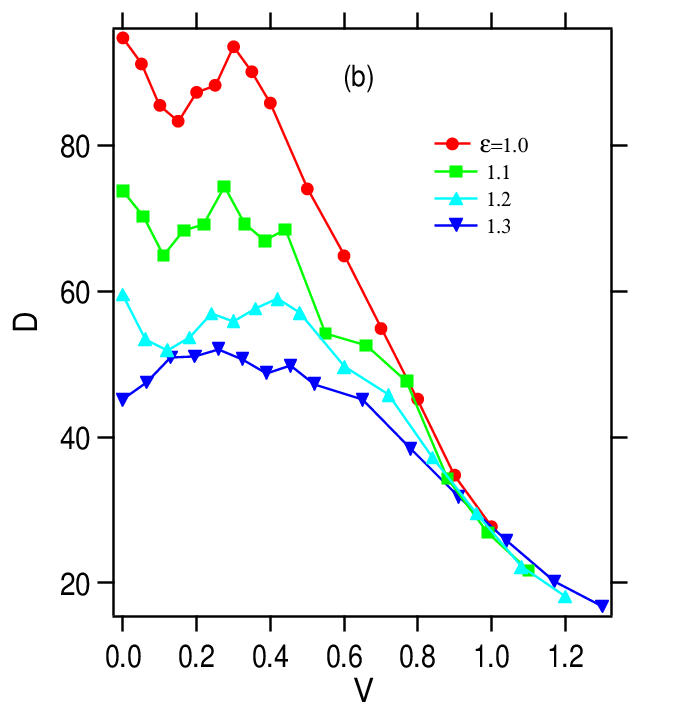}
\caption{(Color online) \label{fig:D-L-3}
(a)The real plots of $m_2$ as a function of $t$ 
for several values of the parameter $0\leq V\leq 1.3$ at the fixed $\eps=1.3$,  
 where the plots of $V=0$ and $V=1.3$ are indicated by the blue and black dotted 
lines, respectively. 
(b)Diffusion coefficient $D$ as a function of $V$  
for various values of $\eps(=1.0,1.1,1.2,1.3)$, where $\hbar=1/8$.
}
\end{center}
\end{figure}

We briefly discuss the $V-$dependence of diffusion coefficients.
Figure \ref{fig:D-L-3}(a) shows the MSD when $V$ is increased
from $V=0$ to $V=1.3$ with fixing 
$\eps=1.3(>>\eps_b\simeq 0.5)$.

At first glance, it may be inferred that the increase in $V$ only
suppresses diffusion, but the change in the diffusion coefficient $D$  
is not simply monotonic with respect to $V$.
Indeed, as can be seen in Fig.\ref{fig:D-L-3}(b),
the change in $D$ has a convex-concave structure 
in the vicinity of $V=0$, which
seems to reflect the fact that $V=0$ is the particular limit implying  the free particle.
But for $V\gtrsim 0.5$ it decreases monotonically.

\section{The bordering region  $V=V_c$} 
\label{sec:V=Vc}
\label{sec:critical-region}

If $\eps$ is large enough the diffusion constant D of $V=V_c$ decays 
following the behaviors of $V>V_c$ and $V<V_c$.  
In this regime the diffusion dynamics continues smoothly to the diffusive behavior of
both sides is continued smoothly is as $V$ varied across $V_c$. 
At least in this regime the diffusive behavior of $V=V_c$ shares 
its nature with those of both sides.

\begin{figure}[htbp]
\begin{center}
\includegraphics[width=6.0cm]{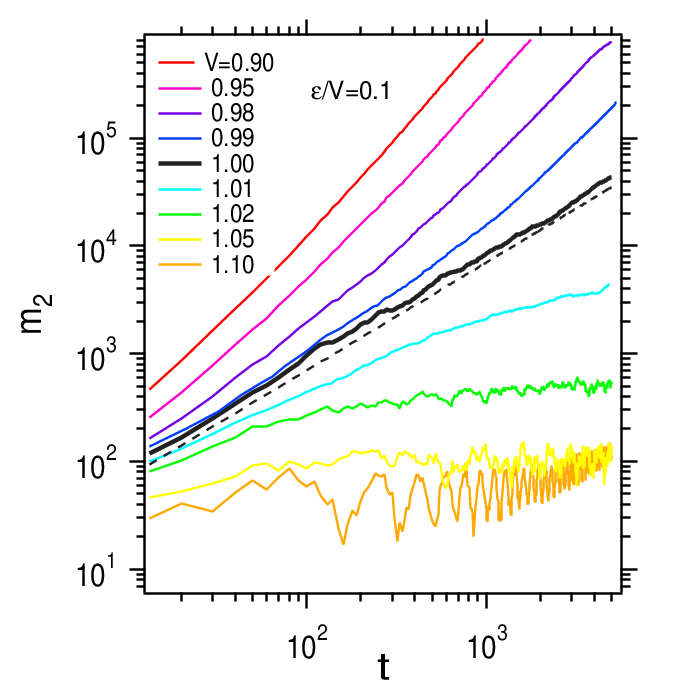}
\caption{(Color online)
\label{fig:c3-e01-v2}
Double-logarithmic plots of $m_2$ as a function of $t$ 
for increasing potential strength $V$ from $V=0$,
where $M=3$ and the relative coupling strength strength  $\eps/V$ is fixed to $0.1$.
The thick solid line denotes the case of $V=1.0$ 
 and the dashed line indicates $m_2 \propto t^1$.
}
\end{center}
\end{figure}

On the other hand at $\eps=0$ the diffive behavior is realized 
as the bordering state between ballistic
state $(V>V_c)$ and localized states $(V>V_c)$, which is due to the duality of Harper model.
Figure \ref{fig:c3-e01-v2} shows the typical behavior of the MSD when $V$ is increased
crossing over the straight line $V=V_c$ with fixing $\eps$ to a sufficiently small value.
It tells that the transition between the localized and ballistic states (LBT)  occurs  
via a diffusive state $m_2\propto t$ at the critical value $V_c=1$ of the unperturbed Harper model.
It just follows the basic feature of the unperturbed Harper model.
This fact implies that at least up to a certain value of $\eps$, the diffusive motion at $V=V_c=1$
remain the same nature as the one due  to the self-duality of Harper model
even if the dynamical perturbation is added.
In fact, as shown in the Appendix \ref{app:BLT-group-velocity},  if $\eps$ is small, 
the group velocity $v_g$ obeys the critical relation $v_g \propto (V_c-V) $ 
in the ballistic side, 
which is a marked character of the unperturbed Harper model ($\eps=0$).

Thus there seemd to exist two regimes of diffusion i.e., $\eps \simeq 0.1$ 
and $\eps\gg 0.5$, along 
the line $V=V_c$ in the $(\eps,V)$-space. 
An anomalous dynamical behavior is observed
between the two regimes.  
Returning to Fig.\ref{fig:unified-D} again, 
we can see that the diffusion coefficient, which is decided by finite time scale
data according to Eq.(\ref{eq:Diffusion}) is not smooth as a function of $\eps$.
This fact implies that the apparently diffusive motion at small $\eps$
described above may temporally accompanied by a large fluctuation on 
a much longer time scale. 

We show in Fig.\ref{fig:critical-msd-long}(a) a long time 
behavior of MSD which corresponds to the short-time MSD data 
 in Appendix \ref{app:V=Vc}. 
The apparently diffusive behavior for the relatively small region, $\eps=0.1 \sim 0.3$,
 fluctuates anomalously  on a long time scale.
In such cases, it is not possible to characterize the motion by a single $D$.
As shown in Fig.\ref{fig:critical-msd-long}(b), 
even the diffusion index $\alpha(t)$ anomalously fluctuates
between $\alpha=0$ and $\alpha=2$, which implies that the mixed
motion among the localized, diffusive, and ballistic motions occurs.

We conjecture that in the anomalously fluctuating
region a transition between the two kinds of normal diffusion, namely the
normal diffusion due to the self-duality of Harper model to the normal diffusion 
induced by the dynamical perturbation, happens in the small $\eps$ region 
indicated by the jugged line in the region in the Fig.\ref{fig:image-fig3}.

We note that the presence of  such an anomalously fluctuating regime
is more pronounced the smaller $M$ is.  (See Appendix \ref{app:V=Vc}.)

\begin{figure}[htbp]
\begin{center}
\includegraphics[width=8.0cm]{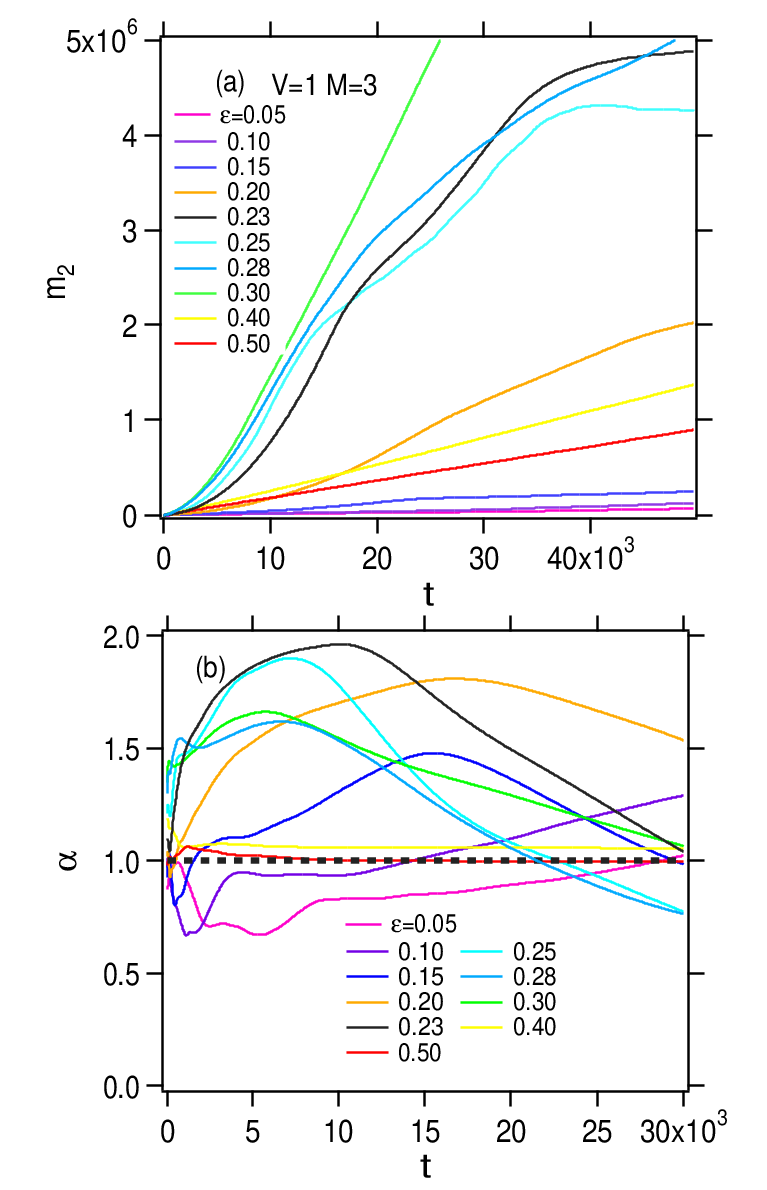}
\caption{(Color online)\label{fig:critical-msd-long}
(a)The time-dependence of $m_2$ on a very long time scale when
the parameters $(V,\eps)$ are taken on the  ``border line''  $V=1=V_{c}$,
where $M=3$.
(b)The time-dependence of $\alpha$ for various strength $\eps$
in the case of $V=1$.
The dotted line indicates $\alpha=1$.
 Note that the both axes are real scale.
}
\end{center}
\end{figure}

\section{Summary and discussion}
\label{sec:summary}
We investigated the quantum wave-packet dynamics of the Harper model
perturbed by  harmonic time-dependent perturbations. We consider this  
model as a typical example that contains neither spatio-temporal randomness
nor singularities such as those in kicked perturbations.

The Harper model has three states-localized state, the ballistic state, and the
critical-depending upon potential strength.
For all these cases, we examined the effect of dynamical perturbation 
by varying the potential strength $V$, the number of colors $M$ and the perturbation 
strength $\eps$. 
For $M \geq 3$, the presence of LDT from the localized side to normal diffusion 
and BDT from the ballistic side to normal diffusion were confirmed.
Such transitions occur even though there is no quasi-periodic static potential 
and the system is ideally ballistic if $\eps=0$.
Further, in the case of the critical state,  there appears to be a transition between 
two types of normal diffusion due to different physical origins, and anomalously
fluctuating diffusion is observed in the transition region.
These results are summarized in the phase diagram shown by Fig.\ref{fig:image-fig3}.
The critical values $\eps_c$ for LDT and and  $\eps_b$ for  BDT  both decreases 
with increasing $M$, and  the region of the normal diffusive phase expands, 
eventually filling most of the $(\eps,V)$-space in the large $M$ limit ($M \ge 3)$.

As a result, a small number of harmonic oscillations can induce a transition to  
an apparently irreversible quantum normal diffusion, despite the the absence of spatial 
and temporal randomness or singularity caused by kicks.
Table \ref{fig:table1} summarizes the localization/delocalization behavior in 
the kicked Harper model (KHM), in addition to the results of this study.

\begin{table}[htbp]
\begin{center}
 \caption{\label{fig:table1}
 $M-$dependence of the DLT and BDT.
For $4 \leq M <\infty$ the result is same as the case of $M=3$.
Loc: exponential localization, Diff:normal diffusion, Balli:ballistic spreading.
}
 \begin{tabular}{lccccc}
\hline
\hline
$M$ & 0 & 1 & 2 & 3 & 4 \\ \hline 
Harper model($V>1$)  & Loc & Loc & Loc & LDT & LDT \\ 
Harper model($0<V<1$)  & Balli & Balli & Balli & BDT & BDT \\ 
KHM ($V>>1$) \cite{yamada23} & Loc & Loc & LDT & LDT  &LDT \\
KHM ($V<<1$) \cite{yamada23} & Balli & Balli & BDT & BDT  &BDT \\
\hline
 \end{tabular}
\end{center}
 \end{table} 



It is worthwhile to explore the dynamical properties under the time-dependent perturbations
in the quasi-periodic models with mobility edges 
\cite{ganeshan15,li17,alexAn21} 
and hierarchical structure of the energy spectrum
 \cite{wilkinson94,thiem09}.
Understanding the robustness of the quantum dynamics induced by 
simple dynamical perturbations
 is a fundamental issue not only for quantum device fabrication, 
 Anderson transitions \cite{abrahams10,tarquini17}, quantum chaos \cite{notarnicola18}, 
 but also for quantum biology, such as the maintenance 
 of quantum coherence at room temperature \cite{vattay14}.

\section*{Acknowledgments}
This work was supported by public funds from Japanese taxpayers through 
MEXT/JSPS KAKENHI Grant Numbers 22K03476 and 22H01146. 
The authors would like to express their sincere gratitude for this support. 
The authors also wish to thank Kankikai (Dr. T. Tsuji) and 
the Koike Memorial House 
for providing access to their facilities during the course of this study.


\appendix

\section{Aubry transform and Self-duality of the Harper model}
\label{app:duality}
In this appendix, we provide a brief explanation of the self-duality in the Harper model 
using the Aubry transform.
We also apply the transform to the system discussed in the main text.

\subsection{Self-duality and localization length}
The time-independent Schrödinger equation of 
the Harper model is given by
\beq
\label{eq:a-eq}
T(a_{n+1}+a_{n-1} )+ 2V\cos(2\pi Qn+\theta)a_n=Ea_n,
\eeq
where $a_n$($n=-\infty,...,\infty$) denotes amplitude at site $n$.
The lattice constant is set to 1, and the wavenumber is $2\pi Q$,  
where $Q$ is an irrational number. In the main text, we take $T=-1$ and $V>0$.
Using the following Aubry's transform (and inverse transform):
\beqnn
  a_n&=&\sum_{m=-\infty}^{\infty}b_m\e^{im(2\pi Qn+\theta)} \e^{i\theta m}, \\
  b_m&=&\sum_{n=-\infty}^{\infty}a_n\e^{-in(2\pi Qm+\theta)} \e^{-i\theta n},
\eeqnn
the expression for the amplitude $b_n$ in reciprocal lattice space
becomes 
\beq
\label{eq:b-eq}
V(b_{m+1}+b_{m-1} )+ 2T\cos(2\pi Qm+\theta)b_m=Eb_m.
\eeq
Compared to the Harper model in Eq.(\ref{eq:a-eq}), 
the roles of $V$ and $T$ is interchanged in Eq.(\ref{eq:b-eq}).
Therefore, $V=T$ is a fixed point of the transform, and 
it exhibits the same energy spectrum.

By applying the  Herbert-Jones-Thouless formula to Eq.(\ref{eq:a-eq}) with $V>|T|$, 
we can obtain the Lyapunov exponent $\gamma_a$, inverse of the localization length: 
\beq
\gamma_a(E)=\int_{-\infty}^\infty \ln \left| \frac{E-E^{`}}{T}\right|dN(E^{`}), 
\eeq
where $N(E^{`})$ is the integrated density of states \cite{sokoloff85}.
Similarly, we obtain the Lyapunov exponent $\gamma_b$ 
for Eq.(\ref{eq:b-eq});
\beq
\gamma_b(E)=\int_{-\infty}^\infty \ln \left|\frac{E-E^{`}}{V} \right|dN(E^{`}).
\eeq
Therefore,
\beq
 \gamma_a(E)=\gamma_b(E)+\ln \left|\frac{V}{T}\right|.
\eeq
If $\gamma_b(E)=0$ and the state is extended in the reciprocal lattice space, 
it is localized in real space as follows:
\beq
 \gamma_a(E)=\ln\left|\frac{V}{T}\right|>0.
\eeq
That is, $V=|T|$ is the transition point, and the localization length is not dependent 
on the energy.

\subsection{Representation in time-dependent systems}
The Aubry transform can also be applied to the model 
with the dynamical perturbation. 
The corresponding time-dependent Schrödinger equations 
for $a_n$ and $b_m$ are given by:
\beq
\label{eq:t-a-eq}
i\hbar \frac{da_n(t)}{dt}&=&T(a_{n+1}(t)+a_{n-1}(t) ) \nn \\
&+&2\cos(2\pi Qn)(V+\eps f(t))a_n(t),
\eeq
\beq
\label{eq:t-b-eq}
i\hbar \frac{db_m(t)}{dt}&=&(V+\eps f(t))(b_{m+1}(t)+b_{m-1}(t) ) \nn \\
&+&2T\cos(2\pi Qm)b_m(t), 
\eeq
where we set $\theta=0$.
In the reciprocal lattice space, the dynamical perturbation appears in the hopping term.
It can be seen that the same dynamical 
phenomena can be described equivalently by both the real-space model 
and its reciprocal-space counterpart.
The  Hamiltonian in the real and 
reciprocal lattice spaces are given respectively by:
\beq
   H_a=2[V+ \eps f(t)] \cos(2\pi Q\hatn)+2T\cos(\hatp/\hbar),
\label{eq:harper-a-H}
\eeq
and 
\beq
   H_b=2T\cos(2\pi Q\hatn)+2[V+\eps f(t)]\cos(\hatp/\hbar), 
\label{eq:harper-b-H}
\eeq
where $\hatn$ and $\hatp$ are the position and momentum operators, respectively.
The Eq.(\ref{eq:harper-a-H}) represents a system in which 
the dynamical perturbation is applied to the on-site potential, as discussed in the main text.
In contrast, Eq.(\ref{eq:harper-b-H}) describes a system 
where the dynamical perturbation acts on the hopping term, this is the dual system.

In the main text, we set $T=-1$ and used Eq. (\ref{eq:t-a-eq}) 
 to investigate LDT for $V>1$ and BDT for $V<1$.
On the other hand, if we use the Eq.(\ref{eq:t-b-eq}),
it corresponds to investigating BDT for $V>1$ and LDT for $V<1$.

\subsection{A numerical result}
Using the case of $M=3$ as an example, 
we examine the dynamical transition to the normal diffusion 
of wave packets in real and reciprocal lattice spaces.

\begin{figure}[htbp]
\begin{center}
\includegraphics[width=4.4cm]{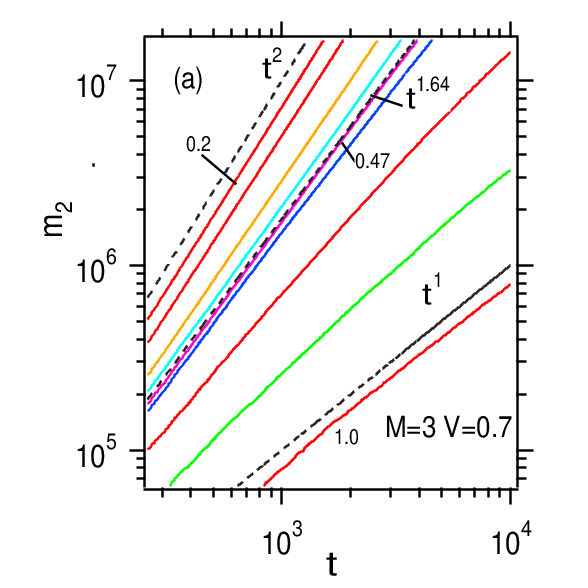}
\hspace{-5mm}
\includegraphics[width=4.4cm]{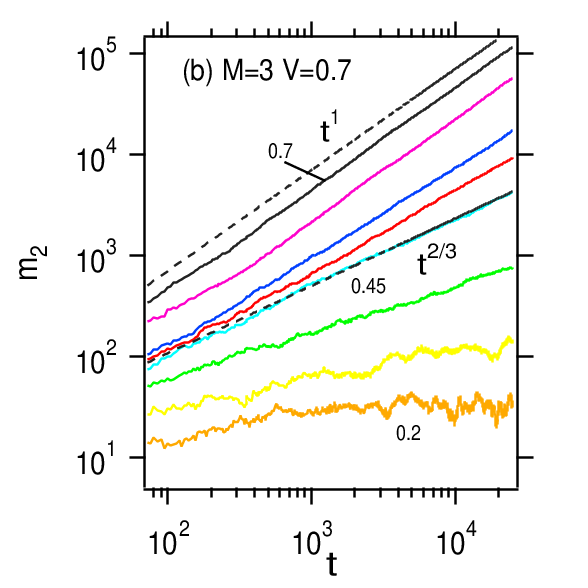}
\caption{(Color online)
\label{fig:c3-msd}
The double logarithmic plots of $m_2$ as a function of $t$ 
for some values of the perturbation strength $\eps/V$
in the  perturbed Harper model ($M=3$) with $V=0.7$. 
(a)BDT when the dynamical perturbation is applied to the on-site term
 given by Eq.(\ref{eq:t-a-eq}).
The dashed lines have slope 2, 1.64,  and 1, respectively. 
(b)LDT when the dynamical perturbation is applied to the hopping term 
given by Eq.(\ref{eq:t-b-eq}).
The dashed lines have slope 2/3 and 1, respectively. 
}
\end{center}
\end{figure}

As seen in Fig.\ref{fig:c3-msd}(a), 
in the case of Eq.(\ref{eq:t-a-eq}) with $V=0.7$ (ballistic regime), 
increasing the strength $\eps$ the BDT causes 
via the superdiffusion as 
\beq
 m_2 \simeq t^{\alpha_b},  \alpha_b \simeq 1.64 
\eeq
at $\eps/V=\eps_b/V$($\simeq 0.45-0.47$).
The corresponding behavior can also be observed  
using Eq.(\ref{eq:t-b-eq}) with $V=0.7$, 
where increasing $\eps$ induces the LDT 
at $\eps/V=\eps_c/V$($\simeq 0.45$), 
as shown in Fig.\ref{fig:c3-msd}(b).
Taking into account a margin of error of $5\%-10\%$, the results are in 
good agreement.

\section{LDT of Maryland model}
\label{app:maryland}
In this appendix, we consider the LDT of Maryland model 
under the dynamical perturbation given by
\beq
\label{eq:maryland}
i\hbar \frac{da_n(t)}{dt}&=&(a_{n+1}(t)+a_{n-1}(t) ) \nn \\
&+&(V+\eps f(t))\tan(\pi Qn+\theta)a_n(t).
\eeq
The unperturbed Maryland model ($\eps=0$) exhibits a singularity 
due to the on-site potential $v(n) = \tan(\pi Q n + \theta)$, 
and it lacks self-duality.
It is so called because the diagonal term has 
the same tangent-type potential 
as in the case of the Maryland transform 
of the kicked rotor systems \cite{yamada20,yamada21,yamada22}.
The eigenstates are always localized with energy-dependent 
localization length, for any finite potential strength $V>0$.
Namely, the spectrum is purely point for $V>0$.

\begin{figure}[htbp]
\begin{center}
\includegraphics[width=4.40cm]{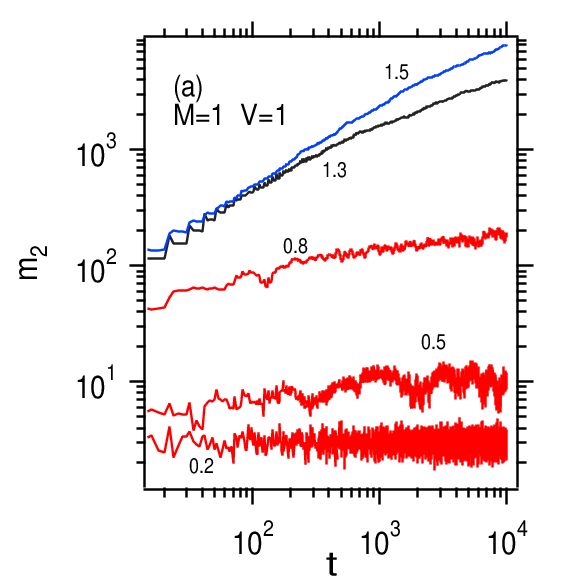}
\hspace{-5mm}
\includegraphics[width=4.40cm]{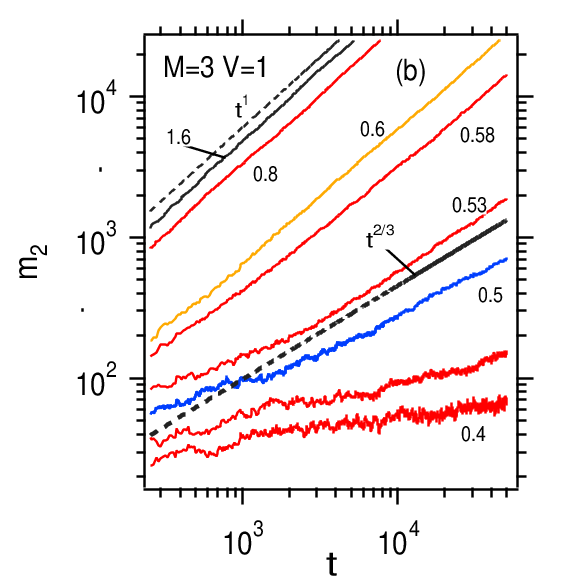}
\caption{(Color online)
\label{fig:maryland-msd}
The double logarithmic plots of $m_2$ as a function of $t$ 
for some values of the perturbation strength $\eps$
in the  perturbed Maryland model with the potential strength $V=1.0$ and $\theta=0$.
 (a)$M=1$ and (b)$M=3$. $\hbar=1/8$.
The dashed lines indicate $m_2 \propto t^1$ and $m_2 \propto t^{2/3}$ in each case.
}
\end{center}
\end{figure}

For mono-chromatically perturbed case, $M=1$, 
the localization is preserved without the transition (absence of transition)
even as the  perturbation strength $\eps$ increases,
as seen in Fig.\ref{fig:maryland-msd}(a).
In the case of $M=3$,  the LDT  appears around $\eps_c\simeq 0.53$,
where $m_2 \sim t^{2/3}$, 
as seen in Fig.\ref{fig:maryland-msd}(b).


\section{Diffusion coefficient for $V=0$}
\label{app:DofV0}
\begin{figure}[htbp]
\begin{center}
\includegraphics[width=5.0cm]{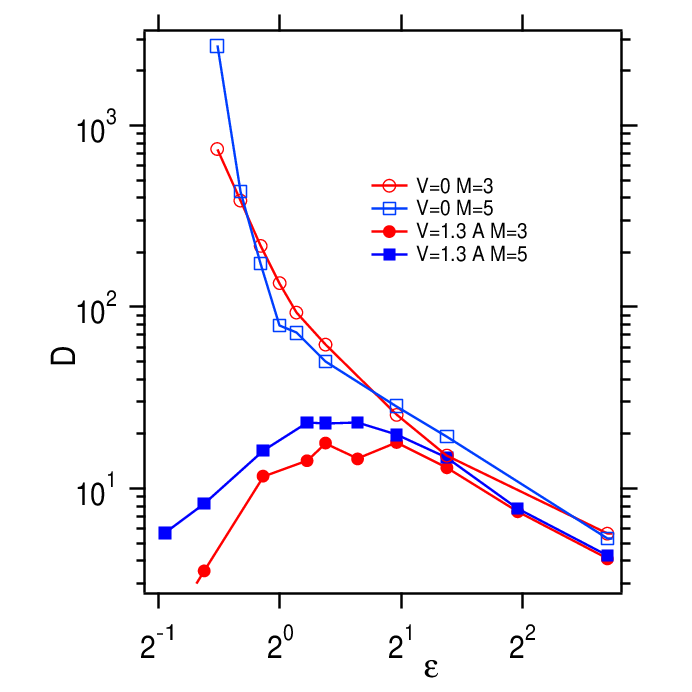}
\caption{\label{fig:tcont-AB-Deps.eps}(Color online)
Diffusion coefficient $D$ as a function of $\eps$ 
for $M=3$ and $M=5$
in the case of $V=0$.
For comparison, the result in the case of $V=1.3$ is also shown.
 Note that the both axes are logarithmic.
}
\end{center}
\end{figure}

The diffusion coefficient $D$ as a function of $\eps$, 
estimated in the normal diffusive region ($\eps>\eps_b$), 
 is shown in Fig.\ref{fig:tcont-AB-Deps.eps} for $V=0$ and $M\geq3$.
It is observed that for the larger $\eps$
 the $D$ decreases monotonically with increasing $\eps$, 
eventually reaching the same level as in the case of $V\ne 0$ when $\eps>>1$.
However, for $\eps<\eps^*$ the decrease in the $D$ deviates significantly 
from the expected scaling $D\sim \eps^{-3/2}$ rule, and the decay is faster 
than in the case of $0<V<V_c$.


\section{Property of the phase B}
\label{app:BLT-group-velocity}
Spreading of the wave packet in the ballistic side  
can be characterized using group velocity $v_g$:
\beq
 v_g^2=\lim_{t \to \infty}\frac{m_2}{t^2}.
\eeq

\begin{figure}[htbp]
\begin{center}
\includegraphics[width=5.0cm]{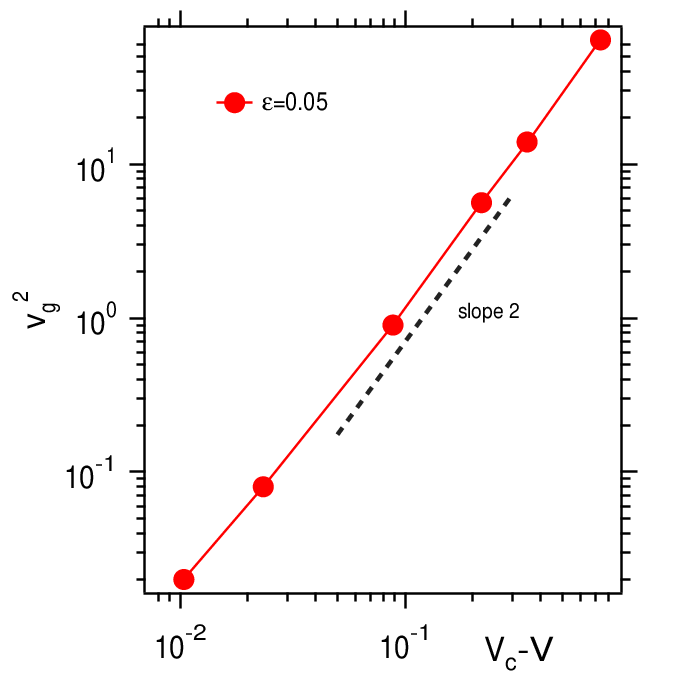}
\caption{(Color online)
\label{fig:Vg-Lc-1.eps}
The squared group verocity $v_g^2$ as a function of $(V_c-V)$
 in the perturbed Harper model with $M=3$ and $\eps=0.05$.
 The numerical data in Fig.\ref{fig:c3-e01-v2} are used.
The dotted line has a slope 2.
}
\end{center}
\end{figure}

The result for $v_g^2$ as a function of  $(V_c-V)$ 
 is shown in Fig.\ref{fig:Vg-Lc-1.eps}.
\beq
 v_g \sim  (V_c-V)^1,
\eeq
 is observed.
This behavior is of the same type as found on the ballistic side ($V <V_c$)
in the (unperturbed) Harper model \cite{rayanov15}.


\section{MSD for the case of $V=V_c$}
\label{app:V=Vc}


We investigate the effect of the perturbation 
 on the critical state of the Harper model $V=V_c(=1)$
 over relatively small time-scale, 
  by varying the various parameters.
 
\begin{figure}[htbp]
\begin{center}
\includegraphics[width=4.4cm]{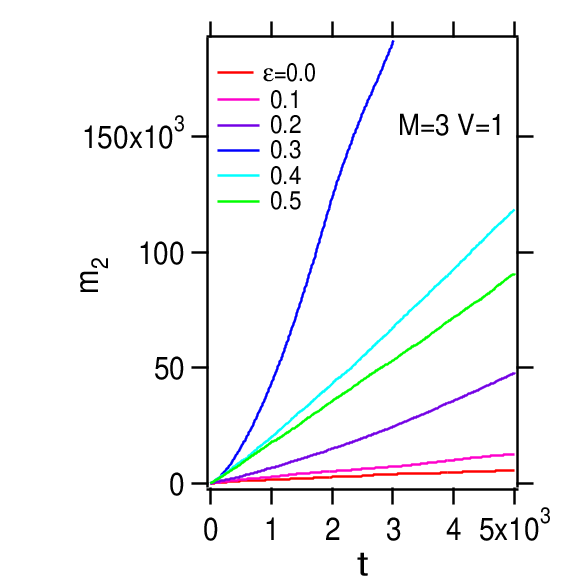}
\hspace{-5mm}
\includegraphics[width=4.4cm]{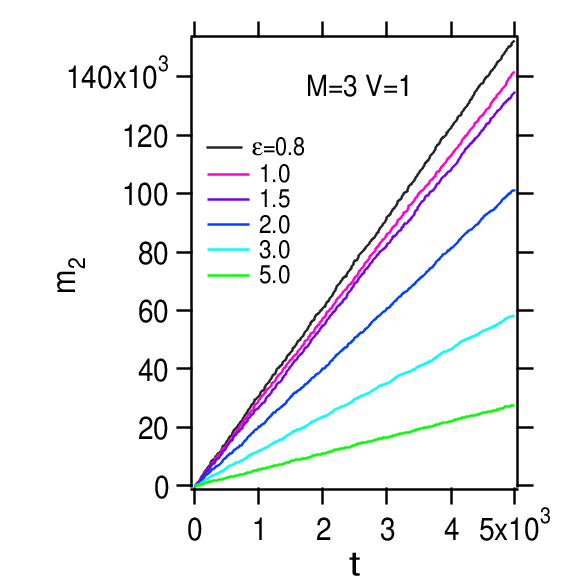}
\hspace{5mm}
\includegraphics[width=4.4cm]{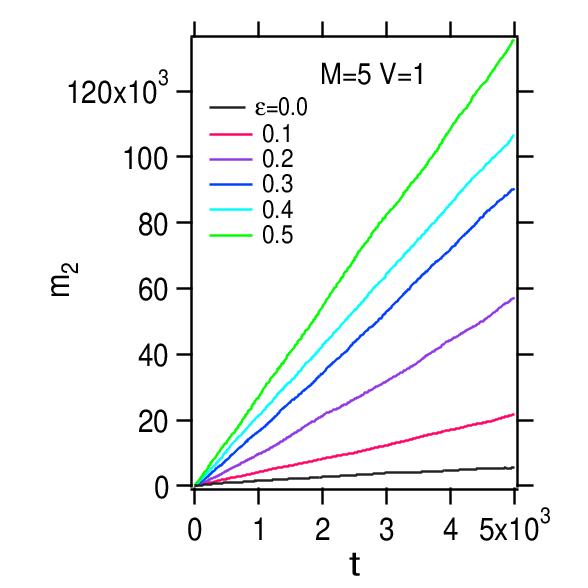}
\hspace{-5mm}
\includegraphics[width=4.4cm]{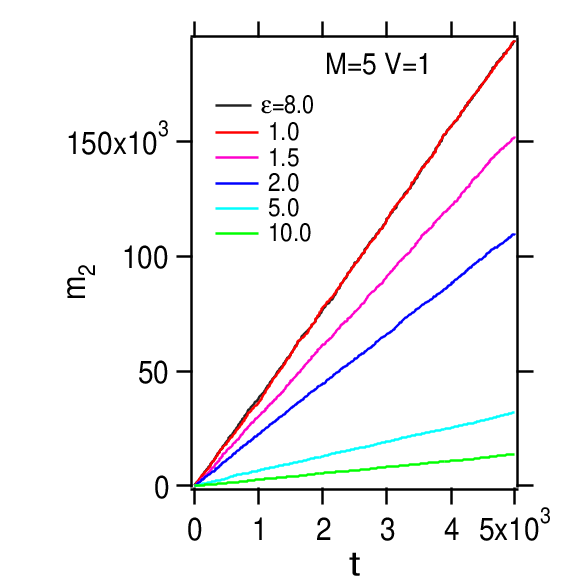}
\caption{(Color online)\label{fig:critical-msd}
The plots of $m_2$ as a function of $t$ 
for some values of $\eps$ in the case of 
the critical potential strength $V=V_c=1.0$.
(a)$M=3$, $\eps \leq0.5$, (b)$M=3$, $\eps \geq0.5$, 
(c)$M=5$, $\eps \leq0.5$, and (d)$M=5$, $\eps \geq0.5$.
 Note that the both axes are real scale.
}
\end{center}
\end{figure}

\begin{figure}[htbp]
\begin{center}
\includegraphics[width=4.4cm]{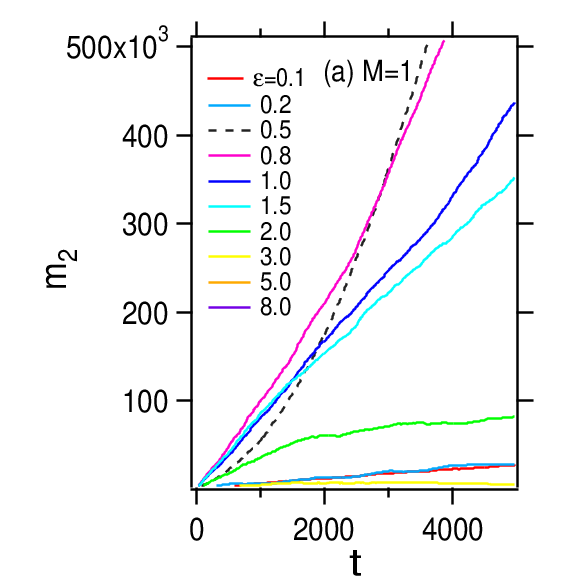}
\hspace{-5mm}
\includegraphics[width=4.4cm]{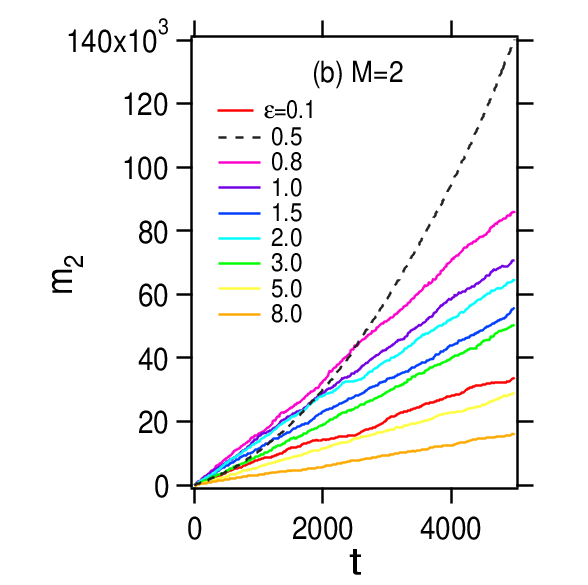}
\caption{(Color online)\label{fig:critical-c1c2-msd}
The plots of $m_2$ as a function of $t$ 
for some values of the perturbation strength $\eps$
in the case of the critical potential strength $V=V_c=1.0$.
(a)$M=1$,  (b)$M=2$.
 Note that the both axes are real scale.
}
\end{center}
\end{figure}

 As shown in Fig.\ref{fig:critical-msd}, the MSDs for different perturbation 
 strength generally exhibit diffusive behavior for $M=3$ and $M=5$.
 In the panels (a) and (c), the amount of diffusion increases with $\eps$,  
 starting from the normal diffusion at $\eps=0$.
 However, in the panels (b) and (d), the amount decreases 
 once $\eps$ exceeds a certain value.
In either case, the normal diffusion at $\eps=0$ will 
eventually returns due to the dynamical perturbation as $\eps$ increases.
A careful look, however, reveals that for the relatively small $\eps(\le 0.2)$, 
$m_2$ also shows a ballistic-like increase.
The complex behavior observed in this region is discussed in Sect.\ref{sec:V=Vc}.
Figure \ref{fig:critical-c1c2-msd} presents the MSD over a wide region of $\eps$
at $V=V_c$ for $M=1$ and $M=2$.
As $M$ decreases, the ballistic growth due to resonance becomes more prominent, 
even at small value of $\eps$. 


\end{document}